\documentclass[twocolumn,pre,twoside,groupedaddress,floatfix,noshowpacs]{revtex4}

\usepackage{amsmath}
\usepackage{amssymb}
\usepackage{graphicx}
\usepackage{bm}
\usepackage{xcolor}

\newcommand{\bs}[1]{\boldsymbol{#1}}

\def\ud{\text{d}}

\begin{document}


\title{Electrolyte solutions at curved electrodes. II. Microscopic approach}

\author{Andreas Reindl}
\email{reindl@is.mpg.de}
\author{Markus Bier}
\email{bier@is.mpg.de}
\author{S. Dietrich}
\affiliation
{
   Max-Planck-Institut f\"ur Intelligente Systeme, 
   Heisenbergstr.\ 3,
   70569 Stuttgart,
   Germany
}
\affiliation
{
   IV. Institut f\"ur Theoretische Physik,
   Universit\"at Stuttgart,
   Pfaffenwaldring 57,
   70569 Stuttgart,
   Germany
}

\date{28 March, 2017}

\begin{abstract}
   Density functional theory is used to describe electrolyte solutions in contact with electrodes of planar or spherical shape.
   For the electrolyte solutions we consider the so-called civilized model, in which all species present
   are treated on equal footing. This allows us
   to discuss the features of the electric double layer in terms of the differential capacitance. The model provides
   insight into the microscopic structure of the electric double layer, which goes beyond the mesoscopic approach
   studied in the accompanying paper.
   This enables us to judge the relevance of microscopic details, such as the radii of the particles
   forming the electrolyte solutions or
   the dipolar character of the solvent particles, and to compare the predictions of various models. 
   Similar to the preceding paper, a general behavior is observed for small radii of the electrode
   in that in this limit the results become independent of the surface charge density
   and of the particle radii. However, for large electrode radii non-trivial behaviors are observed.
   Especially the particle radii and the surface charge density strongly influence the capacitance.
   From the comparison with the Poisson-Boltzmann approach it becomes apparent that the shape of the electrode determines
   whether the microscopic details of the full
   civilized model have to be taken into account or whether already simpler models yield acceptable predictions.
\end{abstract}

\maketitle

\section{\label{section:introduction}Introduction}
In the first part \cite{Reindl2016} of our study electric double layers have been discussed in detail
on mesoscopic scales within the Poisson-Boltzmann (PB) theory,
which was pioneered by Gouy \cite{Gouy1910} and Chapman \cite{Chapman1913}.
The focus of our analysis in Ref.~\cite{Reindl2016} has been on electrodes of spherical or cylindrical shape surrounded by an electrolyte solution.
Such kind of setups serve as models for certain parts of so-called double layer capacitors, i.e.,
electrical energy storage devices which are promising candidates for supporting sustainable energy systems.
In part I \cite{Reindl2016} the differential capacitance $C$, which is the change of the surface charge density upon varying the
electrostatic potential at the wall and thus is experimentally accessible,
has been analyzed in particular concerning its dependence on the surface charge density and on the radius of the curved electrode.
The focus has been, within the PB model, on a thorough discussion of the dependence of the capacitance on the geometry of the system.

Especially for large and small curvatures of the electrodes the behavior has been discussed systematically.
On one hand, the simplifying assumptions underlying the PB approach facilitate such a detailed discussion, and, on the other hand, they are also the reason
for the approach to be only reliable for weak ionic strengths (below $0.2\,\text{M}=0.2\,\text{mol}/\ell$) and low electrode potentials (below $80\,\text{mV}$)
in the case of monovalent salts in aqueous solutions \cite{Butt2003}.
But already in these ranges deviations from experimental data are observable:
the predicted differential capacitance is larger than the measured one \cite{Schmickler2010}.

Therefore Stern introduced a model which accounts for that shortcoming by combining the Gouy-Chapman description of
the diffuse layer, i.e., the charge in the fluid is distributed continuously following the PB equation, with the model of a Helmholtz
layer of counterions, i.e., the charge in the fluid is directly attached to the electrode surface within a molecular layer \cite{Stern1924}.
Accordingly, the pure Gouy-Chapman description, which does not consider the granular character of the fluid, has been improved
by introducing a Stern layer in between the electrode and the diffuse layer, i.e., the actual nonzero particle volumes are
taken into account only close to the wall.
As a consequence the total capacitance of the system can be considered as a circuit of two capacitors
(i.e., the Stern layer and the diffuse layer) in series \cite{Schmickler2010,Butt2003} and, following the rules
for electric circuits, the total capacitance is smaller than the Gouy-Chapman capacitance.
Practically, one is left to fit the capacitance of the Stern layer to experimental results which,
on one hand, might lead to good agreements with measurements
but which, on the other hand, provides only a coarse microscopic picture of the electrical double layer. One possible interpretation
of the results of the Stern theory is that close
to the wall the permittivity is reduced \cite{Butt2003,Schmickler2010}.

More sophisticated models have been developed in order to describe
the structure of an electrical double layer more precisely than within the mesoscopic PB approach
according to which the ions are treated as pointlike charges dissolved in a
homogeneous background. Three classes of corresponding models are common in the literature. Within the
so-called ``primitive model'' (PM) the ions are considered as
charged hard spherical particles which are dissolved 
in a solvent which is taken into account only via the permittivity.
In the so-called ``solvent primitive model'' (SPM) the solvent particles exhibit also a nonzero
volume and are often described as hard spheres.
Sometimes these models are labelled in conjunction with the attribute ``restricted'' which means that all particle radii are equal.
Yet more elaborate theories incorporate the electrostatics between ions and solvent particles by providing the latter with a dipole.

In 1980 Carnie and Chan used the so-called ``civilized model'' in order to model an electrolyte solution \cite{Carnie1980}.
Therein, as opposed to a primitive model, both the ions and the solvent
are treated on an equal basis. The ions are represented by hard spheres with charges whereas the solvent is represented by hard spheres with an embedded dipole.
In Ref.~\cite{Carnie1980} exact and approximative results within the mean spherical approximation are presented for
the structure of the electrolyte solution at a charged planar surface. For low ionic concentrations
the surface potential has the Stern layer form, i.e., the expression of the surface potential is the sum of the diffuse part of the electrical double layer
and an additional part which can be interpreted as the contribution of a Stern layer.
Thus the result is regarded as a derivation of the Stern layer behavior.
Analytic expressions reveal that both the nonzero ion size and the dipolar solvent alter the capacitance,
which otherwise reduces to the double-layer capacitance of the 
linearized Gouy-Chapman theory. However, this change is not very large \cite{Carnie1980}.

In Ref.~\cite{Blum1981} a similar approach is used in order to complement the results of Ref.~\cite{Carnie1980}.
Hard core and solvent effects, both of which are taken into account, lead to a reduction of the differential capacitance, i.e., the trend indicated by the Stern model
is confirmed. However, the  alignment of the dipoles close to the planar wall persists for several layers into the fluid.
This ordering is not confined to a single Stern-like layer next to the electrode.

The so-called reference hypernetted-chain theory is used in Refs.~\cite{Torrie1989,Torrie1989_2} in order to examine the double layer at the surface
of large spherical macroions within a multipolar hard sphere model of electrolyte solutions, i.e., the ions are charged hard spheres and the solvent molecules
are hard spheres carrying point multipoles.
The structure of the double layer is discussed in terms of, e.g., the potential of mean force
between the macroion and a counterion,
ion or solvent number density profiles, and mean electrostatic potentials. The dependences of the structure on the surface charge density,
the ionic concentration, and the (macro-)ionic radii are examined.

In Ref.~\cite{Biben1998} a possible realization of a microscopic model for electric double layers within density functional theory (DFT) is proposed. The functional
is formulated for a mixture of spheres with embedded point charges or dipoles. Consequently, the excess part of the DFT functional splits
into a Coulombic and into hard sphere parts where for the 
latter the fundamental measure theory (see Refs.~[12, 13] in Ref.~\cite{Biben1998}) is adopted.
Within this approach various density profiles near a charged planar wall are presented,
which, due to the granular nature of the fluid, exhibit pronounced oscillations.
The authors conclude by comparison that the primitive model is able to reproduce well ionic density profiles in low concentration regimes and beyond
a certain distance from the wall. However, within the microscopic description and for higher concentrations the
number density oscillations become increasingly pronounced
such that a primitive model description turns out to be inadequate.

Within the microscopic model used in Ref.~\cite{Lamperski1999} the Coulombic interactions are taken into account in a mean-field-like kind
while the excluded-volume effects are accounted for by the Percus Yevick approach. The radii of the various particle species, i.e., anions, cations, and solvent molecules,
are chosen differently. This qualitatively affects the dependence of the differential capacitance $C$ on the surface charge density $\sigma$ of the planar electrode:
within the Gouy-Chapman theory $C$ is an even function of $\sigma$, whereas for unequal values of the particle radii this symmetry disappears.

Oleksy and Hansen have used DFT with an explicit solvent description in order to investigate the wetting and drying behavior
of ionic solutions in contact with a charged solid
substrate \cite{Oleksy2010}. All particles are treated as hard spheres and the corresponding excluded-volume correlations are taken into account by fundamental measure theory.
The other interactions, i.e., the electrostatic interaction and the cohesive Yukawa attraction, are treated within mean field theory in order to ensure full thermodynamic
self-consistency. The key finding is the remarkable agreement between this version of a civilized model and a previous model in which the dipoles are not explicitly taken into account.

Henderson and coworkers proposed a nonprimitive model which differs from the aforementioned models with respect to the shape of the solvent particles \cite{Henderson2012}.
Within DFT hard spherical ions are considered to be dissolved in a solvent which is composed of neutral dimers, i.e., touching positively and negatively charged hard spheres. 
The model predicts a larger electrode potential than in the case of an implicit
model like the restricted primitive model. However, for technical reasons, results could only be obtained for comparatively low potential values.

Recently, electrolyte aqueous solutions near a planar wall have been discussed within a so-called polar-solvation DFT \cite{Warshavsky2016}.
Therein all particles are considered as hard spheres via the fundamental measure theory and the mean spherical approximation is used in order to
calculate the remaining part of the direct correlation functions. The comparison between the results of this polar-solvation DFT,
within which the solvent particles are assumed to carry an
embedded dipole, and the results of the unpolar-solvation DFT of Refs.~\cite{Medasani2014,Ovanesyan2014}, within which
the solvent particles are taken into account as hard spheres only, yields, e.g., a discrepancy in the density profiles which increases with increasing electrode potential.
Due to technical reasons the model in Ref.~\cite{Warshavsky2016} could not be used to study ion concentrations above $10\,\text{mM}$.

For many years a lot of efforts have been spent in order to understand the structure of electric double layers either at
planar walls or around macroions. Therefore various realizations of microscopic
approaches, which in general include an explicit description of dipolar solvents,
have been proposed \cite{Carnie1980,Blum1981,Torrie1989,Torrie1989_2,Biben1998,Lamperski1999,Oleksy2010,Henderson2012,Warshavsky2016}.
Here we study the case of a double layer at a spherical electrode of \textit{arbitrary} radius $R$.
In the spirit of part I \cite{Reindl2016} of this study, the structure of the electrolyte solution
is captured in terms of the differential capacitance $C$, which facilitates to judge the relevance of various system parameters.
As compared to the mesoscopic PB approach used in part I \cite{Reindl2016}, in the present microscopic description the size of the spherical electrode
affects the layering behavior of the particles due to steric effects. In addition, the charges or dipoles, embedded in the particles,
interact with each other and with the charge on the electrode. 
Due to the interplay of various interactions, structural features, such as the spatially varying dipole orientation of the solvent particles,
are expected to exhibit a comparatively complex dependence on the electrode radius.
The model used within the present study is inspired by the work of Oleksy and Hansen \cite{Oleksy2010}.
It incorporates the aforementioned features (non-vanishing and distinct particle volumes as well as spatially varying dipole orientations),
which contribute to the differential capacitance in ways that are,
due to the influence of the electrode size, not yet well understood.
Apart from dealing with the hard spheres the interactions are taken into account
within a random phase description \cite{Evans1979}, the relatively simple
structure of which allows one to conveniently generalize the planar geometry to the spherical one. Moreover, the concept of introducing an
additional cohesive attraction amongst the particles corresponds to our intention to discuss an electrolyte solution in the liquid state under realistic conditions.
To that end the respective model parameters are chosen to mimic liquid water as the solvent at room temperature and ambient pressure. 
The differential capacitance is discussed as function of the remaining inherent system parameters, in particular, the wall curvature.
Here, especially the dependence on the latter
contains mechanisms which are not revealed by more primitive models such that quantitative or even qualitative differences
between our model and the more primitive ones can be expected to occur.

In Sec.~\ref{subsection:density_functional} the present version of the density functional is discussed in detail.
The corresponding Euler-Lagrange equations (ELE) are presented in
Sec.~\ref{subsection:ELE}. The method to account for the boundary conditions in the process of obtaining
the numerical solution is described in Sec.~\ref{subsec:far_from_wall}.
Section~\ref{subsec:parameters} summarizes the chosen parameters and the notation used here.
Technical details are discussed in Appendices~\ref{app:derivation_of_solvent_ELEs}, \ref{app:large_distances}, and \ref{app:c}.
The results of the calculations are presented
in Sec.~\ref{section:discussion} where the structures of the electric double layers are illustrated via spatially varying number density profiles.
Subsequently the capacitance data obtained for the planar wall are shown for various choices of system parameters and models. Finally, the capacitance data of spherical
electrodes are discussed as function of the electrode curvature and of the surface charge density.
The influence of various choices of system parameters is discussed and various models are compared with each other.

\section{\label{section:model}Model}
\subsection{\label{subsection:density_functional}Density functional}
Our microscopic approach follows the one of Oleksy and Hansen in Ref.~\cite{Oleksy2010} which
is sometimes referred to as the so-called \textit{civilized model} \cite{Carnie1980} according to
which both the ions and the solvent are treated on equal footing, in contrast to the primitive model.
We consider an electrolyte solution composed of three species:
solvent particles (hard spheres with radius $r_0$ and an embedded dipole of strength $m$), monovalent cations (hard spheres with radius
$r_1$ carrying a charge $q_1=e>0$, $e$ denoting the absolute value of the elementary charge),
and monovalent anions (hard spheres with radius $r_2$ carrying a charge $q_2=-e<0$).
The spatial extent of the electrolyte solution is restricted due to the presence of an electrode occupying the volume $\mathcal{V}\subseteq\mathbb{R}^3$.
In the present study the focus is on two types of electrodes or walls:
planar electrodes correspond to the half-space $\mathcal{V}=\{(x,y,z)\in\mathbb{R}^3\;|\;z<0\}$ with surface $\mathcal{A}=\{(x,y,z)\in\mathbb{R}^3\;|\;z=0\}$
whereas spherical electrodes of radius $R$
occupy the domain $\mathcal{V}=\{(x,y,z)\in\mathbb{R}^3\;|\;x^2+y^2+z^2<R^2\}$ with surface $\mathcal{A}=\{(x,y,z)\in\mathbb{R}^3\;|\;x^2+y^2+z^2=R^2\}$.
The surfaces $\mathcal{A}$ of the walls may be homogeneously charged
with a surface charge density $\sigma$. The, in general inhomogeneous, distribution of solvent particles is given
by the number density $\varrho_0(\bs{r},\bs{\omega})$, i.e., the number of particles per volume with dipole
orientation $\bs{\omega}$, $|\bs{\omega}|=1$, at position $\bs{r}\in\mathbb{R}^3\backslash\mathcal{V}$.
In a fixed, but otherwise arbitrary, coordinate system the dipole orientation can be represented by
\begin{align}
   \begin{aligned}
      \bs{\omega}=
      \begin{pmatrix}
         \sin(\vartheta)\cos(\varphi)\\
         \sin(\vartheta)\sin(\varphi)\\
         \cos(\vartheta)
      \end{pmatrix}
   \end{aligned}
   \label{eq:omega}
\end{align}
with polar angle $\vartheta$ and azimuthal angle $\varphi$.
The number density of all solvent particles at a point $\bs{r}$ irrespective of the orientation is given by
\begin{align}
   \begin{aligned}
      \bar{\varrho}_0(\bs{r})&:=\int \ud^2\omega\, \varrho_0(\bs{r},\bs{\omega})\\
                             &:=\int\limits_0^\pi \ud\vartheta\,\sin(\vartheta)\int\limits_0^{2\pi}\ud\varphi\,\varrho_0(\bs{r},\vartheta,\varphi).
   \end{aligned}
   \label{eq:rho0_all_dipoles}
\end{align}
In the following the integration over all orientations, i.e., over all possible values of the two angles $\vartheta$ and $\varphi$ [see Eq.~(\ref{eq:omega})],
is denoted as $\int \ud^2\omega$ [see, e.g., Eq.~(\ref{eq:rho0_all_dipoles})].
The number density of ion species $i\in\{1,2\}$ at position $\bs{r}$ is $\varrho_i(\bs{r})$.

Density functional theory (DFT) is a particularly useful approach to determine these density profiles
and with them the structure of the electrolyte solution in contact with the wall \cite{Evans1979}.
To this end we consider the following approximation for the grand potential functional
$\Omega[\varrho_0,\varrho_1,\varrho_2]=:\Omega[\varrho]$
of the number densities $\varrho_0(\bs{r},\bs{\omega})$ and $\varrho_{1,2}(\bs{r})$:
\begin{align}
   \begin{aligned}
      \beta\Omega[\varrho]=&\int \ud^3r\int \ud^2\omega\,\varrho_0(\bs{r},\bs{\omega})\left[\beta V_0(\bs{r})-\beta\mu_0\right]\\
                           &+\sum\limits_{i=1}^2\int \ud^3r\,\varrho_i(\bs{r})\left[\beta V_i(\bs{r})-\beta\mu_i\right]\\
                           &+\beta\mathcal{F}^\text{id}[\varrho]+\beta\mathcal{F}^\text{hs}[\varrho]+\beta\mathcal{F}^\text{el}[\varrho]+
                            \beta\mathcal{F}^\text{att}[\varrho].
   \end{aligned}
   \label{eq:functional}
\end{align}
In Eq.~(\ref{eq:functional}) one has $\beta=(k_BT)^{-1}$ with the Boltzmann constant $k_B$ and the absolute temperature $T$. $V_i$ and $\mu_i$ denote
the external and the chemical potential of species $i$, respectively. The external potential $V_0(\bs{r})$ acting on the solvent particles is taken to be
independent of their dipolar orientations. Unless indicated differently, volume integrals $\int \ud^3r$ run
over the space $\mathbb{R}^3\backslash\mathcal{V}$. $\mathcal{F}^\text{id}$ is the ideal gas contribution,
\begin{align}
   \begin{aligned}
      \beta\mathcal{F}^\text{id}[\varrho]=&\int \ud^3r\int \ud^2\omega\,\varrho_0(\bs{r},\bs{\omega})\left\{\ln\left[\Lambda_0^3\,\varrho_0(\bs{r},\bs{\omega})\right]-1\right\}\\
                                &+\sum\limits_{i=1}^2\int \ud^3r\,\varrho_i(\bs{r})
                                 \left\{\ln\left[\Lambda_i^3\varrho_i(\bs{r})\right]-1\right\},
   \end{aligned}
   \label{eq:Fid}
\end{align}
with the thermal wave lengths $\Lambda_i$.

The hard sphere interaction between the particles is taken into account by means of the functional $\mathcal{F}^\text{hs}$ which, in the present case,
is the White Bear version of the fundamental measure theory (see Ref.~\cite{Roth2002}). For the contribution $\mathcal{F}^\text{hs}$
\{and likewise $\mathcal{F}^\text{att}$ [see Eq.~(\ref{eq:Fatt}) below]\} the orientations $\bs{\omega}$
of the dipoles do not matter which is why it is a functional of $\bar{\varrho}_0(\bs{r})$ [see Eq.~(\ref{eq:rho0_all_dipoles})].

The electrostatic interactions, both amongst the particles and between the particles
and the wall, are captured by the functional
\begin{widetext}
   \begin{align}
      \begin{aligned}
         \beta\mathcal{F}^\text{el}[\varrho]=
              &\frac{1}{2}\sum\limits_{i,j=1}^2\int \ud^3r\int \ud^3r'\frac{\beta}{4\pi\epsilon_0\epsilon_\text{ex}}\frac{q_iq_j}{|\bs{d}|}
                 \varrho_i(\bs{r})\varrho_j(\bs{r'})\Theta[|\bs{d}|-(r_i+r_j)]\\
              &+\sum\limits_{i=1}^2\int \ud^3r \int \ud^3r'\int \ud^2\omega'\frac{\beta}{4\pi\epsilon_0\epsilon_\text{ex}}q_i
                 \frac{m\bs{\omega'}\cdot\bs{d}}{|\bs{d}|^3}\varrho_i(\bs{r})
                 \varrho_0(\bs{r'},\bs{\omega'})\Theta[|\bs{d}|-(r_0+r_i)]\\
              &+\frac{1}{2}\int \ud^3r\int \ud^2\omega\int \ud^3r'\int \ud^2\omega'\frac{\beta}{4\pi\epsilon_0\epsilon_\text{ex}}m^2
                 \left[\frac{\bs{\omega}\cdot\bs{\omega'}}{|\bs{d}|^3}
                 -3\frac{(\bs{\omega}\cdot\bs{d})(\bs{\omega'}\cdot\bs{d})}{|\bs{d}|^5}\right]\varrho_0(\bs{r},\bs{\omega})\varrho_0(\bs{r'},\bs{\omega'})\Theta(|\bs{d}|-2r_0)\\
              &+\sum\limits_{i=1}^2\;\int\limits_{\bs{r}\in\mathcal{A}} \ud^2r\int \ud^3r'\frac{\beta}{4\pi\epsilon_0\epsilon_\text{ex}}\frac{\sigma q_i}{|\bs{d}|}\varrho_i(\bs{r'})
                 \Theta(|\bs{d}|-r_i)\\
              &+\int\limits_{\bs{r}\in\mathcal{A}} \ud^2r\int \ud^3r'\int \ud^2\omega'\frac{\beta}{4\pi\epsilon_0\epsilon_\text{ex}}\sigma\frac{m\bs{\omega'}\cdot\bs{d}}{|\bs{d}|^3}
                 \varrho_0(\bs{r'},\bs{\omega'})\Theta(|\bs{d}|-r_0),
      \end{aligned}
      \label{eq:Fel}
   \end{align}
\end{widetext}
where $\epsilon_0$ is the vacuum permittivity,
\begin{align}
   \begin{aligned}
      \Theta(x)=\begin{cases}
         0,&x<0\\
         1,&x>0
      \end{cases}
   \end{aligned}
   \label{eq:Heaviside}
\end{align}
denotes the Heaviside step function, and $\bs{d}:=\bs{r}-\bs{r'}$ is the spatial offset between positions $\bs{r}$ and $\bs{r'}$.
Within the present approach, the dielectric properties of the solvent are described in terms of an explicit mean-field-like contribution due to the solvent
particles with embedded dipole moments $\bs{p}:=m\bs{\omega}$ and an implicit excess contribution given by the excess relative permittivity
$\epsilon_\text{ex}$ which captures dielectric properties beyond
the mean-field description [see also Sec.~\ref{subsec:far_from_wall} and in particular Eq.~(\ref{eq:epsex})].
Alternatively, one could use descriptions based on the mean spherical approximation \cite{Warshavsky2016}
or model the solvent molecules as dimers in the style of Ref.~\cite{Henderson2012}.
However, the present approach has technical advantages and, moreover, it has turned out that the explicit dipolar contribution affects the results only weakly.

The influence of Coulomb correlation contributions has been examined in Ref.~\cite{Oleksy2006},
where semi-primitive model electrolytes have been described both in terms of a mean-field density functional, which neglects Coulomb correlations, 
and in terms of a more complex density functional including such correlations. The outcome of these different approaches has been compared with each other and with
Monte Carlo simulations. The authors have found that Coulomb correlation corrections alter the mean-field results significantly only for high surface charges
in the presence of divalent cations. Based on that finding, the model in the present study is expected to accurately describe electrolytes consisting of monovalent ions.
In addition, we have compared the results of the present model for an ionic strength of $0.1\,\text{M}$
with Figs.~5 and 6 in Ref.~\cite{Patra2009}, which provide simulation results for a charged
spherical macroparticle surrounded by an electrolyte solution within the molecular solvent model. All profiles show good agreement with the simulation data which supports
the aforementioned argument that the present model is a reliable description for monovalent ions.
Another consistency check has been carried out with respect to the bulk limit. To that end number density profiles
around a spherical electrode, with the same size and charge as an ion, have been calculated. These density profiles, which correspond to pair distribution functions, have been compared
with Fig.~2 in Ref.~\cite{Wu1999} where simulation data for an electrolyte within a solvent primitive model are presented.
Our results almost lie on top of the curves denoted by ``HNC'' and are in good agreement with the simulation data.

An attractive interaction between the particles is taken into account by the contribution
$\mathcal{F}^\text{att}$. This interaction is rationalized by a van der Waals attraction which enables the formation of a liquid state under realistic conditions,
i.e., room temperature and ambient pressure.
Amplitude and range of this potential, chosen to be square-well-like, are given by $u$ and $r_c$, respectively:
\begin{align}
   \begin{aligned}
      \beta\mathcal{F}^\text{att}[\varrho]=\frac{1}{2}\sum\limits_{i,j=0}^2\int \ud^3r\int \ud^3r'\,\beta u\Theta(r_c-|\bs{d}|)\\
                 \times\varrho_i(\bs{r})\varrho_j(\bs{r'})\Theta[|\bs{d}|-(r_i+r_j)].
   \end{aligned}
   \label{eq:Fatt}
\end{align}
[Note that in Eq.~(\ref{eq:Fatt}) $\varrho_0(\bs{r},\bs{\omega})$ reduces to $\bar{\varrho}_0(\bs{r})$].
We make the simplifying assumption that this kind of
attractive potential is the same for the interaction among the fluid particles and for their interaction with the wall particles.
Thus for a homogeneous number density $\varrho_w$ of the wall particles, the attractive van der Waals interaction gives rise to the substrate potential
\begin{align}
   \begin{aligned}
      \beta V^\text{att}(\bs{r})=\int\limits_\mathcal{V}\ud^3r'\,\varrho_w\beta u\Theta(r_c-|\bs{d}|).
   \end{aligned}
   \label{eq:Vatt}
\end{align}
The attractive van der Waals potential $V^\text{att}$ together with
the hard repulsive interaction between wall and fluid particles comprise the external potentials $V_i$, $i\in\{0,1,2\}$:
\begin{align}
   \begin{aligned}
      \beta V_i(\bs{r})=\begin{cases}
         \infty,&\exists\;\bs{r'}\in\mathcal{V}:|\bs{d}|<r_i,\\
         \beta V^\text{att}(\bs{r}),&\text{otherwise}.
      \end{cases}
   \end{aligned}
   \label{eq:V}
\end{align}
Further contributions, e.g., due to electrostatic image forces in the case of a dielectric contrast between wall and fluid, would have to be added
to the second line of Eq.~(\ref{eq:V}).

\subsection{\label{subsection:ELE}Euler-Lagrange equations}
In accordance with the variational principle underlying density functional theory \cite{Evans1979} the equilibrium densities $\varrho^\text{eq}_i$ minimize
the functional in Eq.~(\ref{eq:functional}) and thus fulfill the Euler-Lagrange equations (ELE)
\begin{align}
   \frac{\delta(\beta\Omega)}{\delta\varrho_0(\bs{r},\bs{\omega})}\bigg|_{\varrho^\text{eq}_0,\varrho^\text{eq}_1,\varrho^\text{eq}_2}=0,\quad
      \frac{\delta(\beta\Omega)}{\delta\varrho_{1,2}(\bs{r})}\bigg|_{\varrho^\text{eq}_0,\varrho^\text{eq}_1,\varrho^\text{eq}_2}=0.
   \label{eq:ELE0}
\end{align}
Their forms will be discussed in more detail below (see also Ref.~\cite{Oleksy2010}).
For clarity the superscript $^\text{eq}$ is omitted and subsequently the focus is on equilibrium densities.
In the bulk and thus in the absence of inhomogeneities, the density profiles entering the ELE~(\ref{eq:ELE0}) are uniform and isotropic:
$\varrho_0(\bs{r},\bs{\omega})=\varrho_0^\text{b}/(4\pi)$, $\varrho_1(\bs{r})=\varrho_1^\text{b}\equiv I$, and $\varrho_2(\bs{r})=\varrho_2^\text{b}\equiv I$
with the ionic strength $I$. Note that $\varrho_1^\text{b}=\varrho_2^\text{b}$ implies local charge neutrality.
This also holds at points sufficiently far away from the wall.
By subtracting from the ELE~(\ref{eq:ELE0}) the respective expressions in the bulk, the chemical potentials $\mu_i$ and the lengths $\Lambda_i$
drop out of the equations:
\begin{align}
   \begin{aligned}
      \varrho_0&(\bs{r},\bs{\omega})=\frac{\varrho_0^\text{b}}{4\pi}\exp\Big\{-\beta V_0(\bs{r})+c_0^\text{hs}(\bs{r})-c_0^\text{hs,b}\\
                  &+c_0^\text{att}(\bs{r})-c_0^\text{att,b}+\beta m\bs{\omega}\cdot\left[\bs{E}(\bs{r})-\bs{E}^\text{aux}(\bs{r})\right]\Big\}
   \end{aligned}
   \label{eq:ELE1a}
\end{align}
and, $i\in\{1,2\}$,
\begin{align}
   \begin{aligned}
      \varrho_i&(\bs{r})=\varrho_i^\text{b}\exp\Big\{-\beta V_i(\bs{r})+c_i^\text{hs}(\bs{r})-c_i^\text{hs,b}\\
                  &+c_i^\text{att}(\bs{r})-c_i^\text{att,b}-\beta q_i\left[\Phi(\bs{r})-\Phi_i^\text{aux}(\bs{r})\right]\Big\}.
   \end{aligned}
   \label{eq:ELE1b}
\end{align}
Here the following quantities have been introduced:
one-point direct correlation functions, $\text{x}\in\{\text{hs},\text{att}\}$,
\begin{align}
   c_i^\text{x}(\bs{r}):=-\frac{\delta(\beta\mathcal{F}^\text{x})}{\delta\varrho_i(\bs{r})},\quad
   c_i^\text{x,b}:=-\frac{\delta(\beta\mathcal{F}^\text{x})}{\delta\varrho_i(\bs{r})}\bigg|_\text{bulk},
   \label{eq:c}
\end{align}
the polarization
\begin{align}
   \begin{aligned}
      \bs{P}(\bs{r}):=\int \ud^2\omega\, m\bs{\omega}\varrho_0(\bs{r},\bs{\omega}),
   \end{aligned}
   \label{eq:P}
\end{align}
electric fields
\begin{align}
   \begin{aligned}
      &\bs{E}(\bs{r}):=\sum\limits_{i=1}^2\int \ud^3r'\frac{q_i}{4\pi\epsilon_0\epsilon_\text{ex}}\frac{\bs{d}}{|\bs{d}|^3}\left[\varrho_i(\bs{r'})-\varrho_i^\text{b}\right]\\
      &\quad\quad -\int \ud^3r'\frac{1}{4\pi\epsilon_0\epsilon_\text{ex}}\left\{\frac{\bs{P}(\bs{r'})}{|\bs{d}|^3}-\frac{3\bs{d}[\bs{P}(\bs{r'})\cdot\bs{d}]}{|\bs{d}|^5}\right\}\\
      &\quad\quad +\int\limits_{\bs{r'}\in\mathcal{A}} \ud^2r'\frac{\sigma}{4\pi\epsilon_0\epsilon_\text{ex}}\frac{\bs{d}}{|\bs{d}|^3},
   \end{aligned}
   \label{eq:E}
\end{align}
and
\begin{align}
   \begin{aligned}
      &\bs{E}^\text{aux}(\bs{r}):=\\
      &\sum\limits_{i=1}^2\int \ud^3r'\frac{q_i}{4\pi\epsilon_0\epsilon_\text{ex}}\frac{\bs{d}}{|\bs{d}|^3}\left[\varrho_i(\bs{r'})-\varrho_i^\text{b}\right]\Theta(r_0+r_i-|\bs{d}|)\\
      &-\int \ud^3r'\frac{1}{4\pi\epsilon_0\epsilon_\text{ex}}\left\{\frac{\bs{P}(\bs{r'})}{|\bs{d}|^3}-\frac{3\bs{d}[\bs{P}(\bs{r'})\cdot\bs{d}]}{|\bs{d}|^5}\right\}\\
       &\quad\quad\quad\times\Theta(2r_0-|\bs{d}|)\\
      &+\int\limits_{\bs{r'}\in\mathcal{A}} \ud^2r'\frac{\sigma}{4\pi\epsilon_0\epsilon_\text{ex}}\frac{\bs{d}}{|\bs{d}|^3}\Theta(r_0-|\bs{d}|),
   \end{aligned}
   \label{eq:Eaux}
\end{align}
as well as electric potentials
\begin{align}
   \begin{aligned}
      &\Phi(\bs{r}):=\sum_{j=1}^2\int \ud^3r'\frac{1}{4\pi\epsilon_0\epsilon_\text{ex}}\frac{q_j}{|\bs{d}|}\left[\varrho_j(\bs{r'})-\varrho_j^\text{b}\right]\\
      &\quad+\int \ud^3r'\frac{1}{4\pi\epsilon_0\epsilon_\text{ex}}\frac{\bs{P}(\bs{r'})\cdot\bs{d}}{|\bs{d}|^3}+\int\limits_{\bs{r'}\in\mathcal{A}}\ud^2r'
               \frac{1}{4\pi\epsilon_0\epsilon_\text{ex}}\frac{\sigma}{|\bs{d}|}
   \end{aligned}
   \label{eq:Phi}
\end{align}
and
\begin{align}
   \begin{aligned}
      &\Phi_i^\text{aux}(\bs{r}):=\\
      &\sum_{j=1}^2\int \ud^3r'\frac{1}{4\pi\epsilon_0\epsilon_\text{ex}}\frac{q_j}{|\bs{d}|}\left[\varrho_j(\bs{r'})-\varrho_j^\text{b}\right]\Theta(r_i+r_j-|\bs{d}|)\\
      &+\int \ud^3r'\frac{1}{4\pi\epsilon_0\epsilon_\text{ex}}\frac{\bs{P}(\bs{r'})\cdot\bs{d}}{|\bs{d}|^3}\Theta(r_0+r_i-|\bs{d}|)\\
      &+\int\limits_{\bs{r'}\in\mathcal{A}}\ud^2r'\frac{1}{4\pi\epsilon_0\epsilon_\text{ex}}\frac{\sigma}{|\bs{d}|}\Theta(r_i-|\bs{d}|).
   \end{aligned}
   \label{eq:Phiaux}
\end{align}
The original Heaviside functions inherited from the functional $\mathcal{F}^\text{el}$ in Eq.~(\ref{eq:Fel}) are split according to $\Theta(x)=1-\Theta(-x)$.
This is the reason for the appearance of the auxiliary quantities, denoted with the superscript $^\text{aux}$ [see Eqs.~(\ref{eq:Eaux}) and (\ref{eq:Phiaux})].
Since these quantities act like electric fields or potentials, respectively, we use the same corresponding notions for them.
The advantage of this separation is of technical nature:
on one hand, the integration domains in
$\bs{E}^\text{aux}$ [Eq.~(\ref{eq:Eaux})] and $\Phi_i^\text{aux}$ [Eq.~(\ref{eq:Phiaux})] are bounded and therefore numerically manageable.
On the other hand, the \textit{total} electric field $\bs{E}=-\bs{\nabla}\Phi$ [Eq.~(\ref{eq:E})]
and the \textit{total} electric potential $\Phi$ [Eq.~(\ref{eq:Phi})] are determined by
electrostatics and fulfill Poisson's equation with Neumann boundary conditions:
\begin{align}
   \begin{aligned}
      \Delta\Phi(\bs{r})&=-\frac{1}{\epsilon_0\epsilon_\text{ex}}\sum\limits_{j=1}^2 q_j\varrho_j(\bs{r})+\frac{1}{\epsilon_0\epsilon_\text{ex}}\bs{\nabla}\cdot\bs{P}(\bs{r}),\\
      &\int\limits_{\bs{r}\in\mathcal{A}} \ud^2r\,\bs{n}(\bs{r})\cdot\bs{E}(\bs{r})=\frac{\sigma}{\epsilon_0\epsilon_\text{ex}}|\mathcal{A}|,\\
      &\lim\limits_{\lambda\rightarrow\infty}\int\limits_{\bs{r}\in\mathcal{B}(\lambda)} \ud^2r\,\bs{n}(\bs{r})\cdot\bs{E}(\bs{r})=0,\\
      &\mathcal{B}(\lambda):=\left\{\bs{r}+\lambda\bs{n}(\bs{r})\in\mathbb{R}^3|\bs{r}\in\mathcal{A}\right\}.
   \end{aligned}
   \label{eq:Poisson}
\end{align}
Hence, $\bs{E}$ and $\Phi$ are the solution of the boundary value problem posed by Eq.~(\ref{eq:Poisson}).
Since the differential equation~(\ref{eq:Poisson}) has to be evaluated only locally, this route is technically more convenient than 
to perform the integrals over the whole space in Eqs.~(\ref{eq:E}) and (\ref{eq:Phi}).
The unit vectors $\bs{n}$ in the second and in the last line of Eq.~(\ref{eq:Poisson}) point into the radial direction away from the wall
and are locally perpendicular to the respective surface. $|\mathcal{A}|$ denotes the area of the wall surface $\mathcal{A}$. Poisson's
equation~(\ref{eq:Poisson}) determines the potential $\Phi$ up to an additive constant which is chosen such that $\lim\limits_{\bs{r}\rightarrow\infty}\Phi(\bs{r})=0$, i.e.,
the potential $\Phi(\bs{r})$ at any position $\bs{r}$ corresponds to the voltage with respect to the bulk at large distances from the wall $(\bs{r}\rightarrow\infty)$.
We note that the equations for the density profiles [see Eqs.~(\ref{eq:ELE1a})
and (\ref{eq:ELE1b})] depend only on the differences $\bs{E}-\bs{E}^\text{aux}$ and $\Phi-\Phi_i^\text{aux}$.

Due to the dependence of the ELE~(\ref{eq:ELE1a}) on both the position $\bs{r}$ \textit{and} the orientation $\bs{\omega}$, in general the problem has to be
solved in a high-dimensional space which is difficult to handle. However, for certain geometries of the electrode, this dimension can be reduced to a large
extent \cite{Oleksy2010}. To this end the distribution function $f$ of the dipole orientations is introduced and expanded in terms of spherical harmonics:
\begin{align}
   f(\bs{r},\bs{\omega}):=\frac{\varrho_0(\bs{r},\bs{\omega})}{\bar{\varrho}_0(\bs{r})}=\sum\limits_{l=0}^\infty\sum\limits_{m=-l}^lf_{lm}(\bs{r})Y_{lm}(\vartheta,\varphi).
   \label{eq:f}
\end{align}
Due to the definition of the orientation independent number density $\bar{\varrho}_0(\bs{r})$ in Eq.~(\ref{eq:rho0_all_dipoles}), $f$ is normalized, i.e.,
\begin{align}
   \int \ud^2\omega\, f(\bs{r},\bs{\omega})=1,
\end{align}
which determines the value of the coefficient $f_{00}(\bs{r})=(4\pi)^{-1/2}$. In principle, the expansion in Eq.~(\ref{eq:f}) leads to a dependence of
the polarization $\bs{P}$ [Eq.~(\ref{eq:P})]
on the coefficients $f_{lm}$ of order $l=1$, i.e., on $f_{1,0},f_{1,1}$, and $f_{1,-1}$. However, for planar and spherical electrodes
the orientation of $\bs{P}$ is perpendicular to the electrode surface everywhere and to all corresponding parallel surfaces.
The respective normal component $P$ is
\begin{align}
   P(\bs{r})=\sqrt{\frac{4\pi}{3}}m\bar{\varrho}_0(\bs{r})f_{1,0}(\bs{r}),
   \label{eq:Pcomp}
\end{align}
i.e., the polarization depends only on the coefficient $f_{1,0}$, provided that the polar axis of the spherical harmonics is chosen perpendicular to
the electrode surface pointing away from the wall.
Likewise the electric fields $\bs{E}$ and $\bs{E}^\text{aux}$ are
perpendicular to the surface $\mathcal{A}$ with components $E$ and $E^\text{aux}$, respectively.
This facilitates integration over the orientations $\bs{\omega}$ such that
the ELE~(\ref{eq:ELE1a}) for $\varrho_0(\bs{r},\bs{\omega})$ can be split into two equations (see Appendix \ref{app:derivation_of_solvent_ELEs}):
one for the orientation independent density
\begin{align}
   \begin{aligned}
      \bar{\varrho}_0&(\bs{r})=\varrho_0^\text{b}\Big\{\exp\Big[-\beta V_0(\bs{r})+c_0^\text{hs}(\bs{r})-c_0^\text{hs,b}\\
         &+c_0^\text{att}(\bs{r})-c_0^\text{att,b}\Big]\Big\}\frac{\sinh\{\beta m[E(\bs{r})-E^\text{aux}(\bs{r})]\}}{\beta m[E(\bs{r})-E^\text{aux}(\bs{r})]},
   \end{aligned}
   \label{eq:ELE_rho0}
\end{align}
and another one for the coefficient
\begin{align}
   \begin{aligned}
      f_{1,0}(\bs{r})=\sqrt{\frac{3}{4\pi}}\mathcal{L}\{\beta m[E(\bs{r})-E^\text{aux}(\bs{r})]\}
   \end{aligned}
   \label{eq:ELE_f10}
\end{align}
with the Langevin function $\mathcal{L}(x)=\coth(x)-1/x$. Altogether the model is described by four equations:
one for each of the number densities $\varrho_i(\bs{r})$ of the ion species $i\in\{1,2\}$ 
[see Eq.~(\ref{eq:ELE1b})], one for the solvent number density $\bar{\varrho}_0(\bs{r})$ independent of the dipole orientation [see Eq.~(\ref{eq:ELE_rho0})],
and one for the orientation coefficient $f_{1,0}(\bs{r})$ of the dipoles [see Eq.~(\ref{eq:ELE_f10})]. The entire dependence on the orientation $\bs{\omega}$
is covered by the latter quantity. Therefore and because in the case of planar and spherical electrodes the four profiles $\bar{\varrho}_0$, $\varrho_1$,
$\varrho_2$, and $f_{1,0}$ vary only along the direction perpendicular to the surface, the dimensionality of the original problem has been reduced considerably.

\subsection{\label{subsec:far_from_wall}Behavior at large distances from the wall}
Equations~(\ref{eq:ELE1b}), (\ref{eq:Eaux}), (\ref{eq:Phiaux}), (\ref{eq:Poisson}), (\ref{eq:Pcomp}), (\ref{eq:ELE_rho0}),
and (\ref{eq:ELE_f10}) form a complicated
system of coupled nonlinear integro-differential equations, which can be
solved only numerically.
This requires discretization of the various profiles on a large but finite
grid along the radial direction.
This approach requires assumptions concerning the profiles outside the numerical grid
at large distances from the wall.
Here it is assumed that the one-point direct correlation functions 
$c_i^\text{hs}(\bs{r})$ and $c_i^\text{att}(\bs{r})$ decay rapidly
towards their bulk values $c_i^\text{hs,b}$ and $c_i^\text{att,b}$,
respectively, such that in Eqs.~(\ref{eq:ELE1a}) and (\ref{eq:ELE1b}) the differences $c_i^\text{hs}(\bs{r})-
c_i^\text{hs,b}$ and $c_i^\text{att}(\bs{r})-c_i^\text{att,b}$
can be neglected outside the numerical grid. (See Appendix~\ref{app:c} for a detailed discussion of the decay behavior of these one-point
direct correlation functions.)
Global charge neutrality requires the vanishing of the electric field $\bs{E}$
infinitely far away from the wall [see the third line of Eq.~(\ref{eq:Poisson})]. 
Since a numerical grid can span only a finite distance from the wall, the solution of Poisson's equation~(\ref{eq:Poisson}) in the
asymptotic range outside the grid has to be determined, e.g., in terms of a linearized theory and matched with the numerical solution inside the grid.
Therefore the electric fields $\bs{E}(\bs{r})$ and $\bs{E}^\text{aux}(\bs{r})$ as well as the
potentials $\Phi(\bs{r})$ and $\Phi_i^\text{aux}(\bs{r})$ are not required to
vanish outside the numerical grid. Instead it is assumed that at large distances from the wall the electric fields and potentials 
are sufficiently small to allow for a linearization of the exponential function in Eqs.~(\ref{eq:ELE1a}) and (\ref{eq:ELE1b}) such that
the ELEs are given by Eqs.~(\ref{eq:ELE1a_approx}) and (\ref{eq:ELE1b_approx}) in Appendix~\ref{app:large_distances}.
It can be shown by means of Eqs.~(\ref{eq:E})--(\ref{eq:Phiaux}) that for equally-sized ions and weakly charged walls
the quantities $E_x-E_x^\text{aux}$, $E_y-E_y^\text{aux}$, 
$E_z-E_z^\text{aux}$, $\Phi-\Phi_1^\text{aux}$, and
$\Phi-\Phi_2^\text{aux}$ (i) exhibit the same asymptotic decay behavior at large distances from the wall $(\bs{r}\rightarrow\infty)$
and (ii) are proportional to the surface charge density (see Appendix \ref{app:large_distances} for details).
We make use of this property by introducing \textit{constants} $k^E$, $k_1^\Phi$, and $k_2^\Phi$,
\begin{align}
   k_i^\Phi:=-\lim\limits_{\bs{r}\rightarrow\infty}\frac{\Phi_i^\text{aux}(\bs{r})}{\Phi(\bs{r})},\quad k^E:=-\lim\limits_{\bs{r}\rightarrow\infty}\frac{E^\text{aux}(\bs{r})}{E(\bs{r})},
   \label{eq:constants}
\end{align}
for systems with weakly charged planar walls,
where $E$ and $E^\text{aux}$ denote the radial components of the electric fields.
Since both numerators and denominators in Eq.~(\ref{eq:constants}) exhibit the same decay behavior,
as well as the
asymptotic proportionality to $\sigma$ in the limit $\sigma\rightarrow0$, to leading order
the constants do not vary spatially and do not depend on the surface charge density.
It has turned out numerically that the constants $k^E$, $k_1^\Phi$, and $k_2^\Phi$ as
determined for a weakly charged planar wall are valid for all curvatures and surface charges used in the present study.
Moreover, we have found that this procedure works also for small differences between the particle radii, which are
at most as large as the ones considered in the following.
By using Eq.~(\ref{eq:constants}) the asymptotically leading contribution to the auxiliary fields in Eqs.~(\ref{eq:ELE1a_approx}) and (\ref{eq:ELE1b_approx})
in Appendix~\ref{app:large_distances} can be expressed in terms of the constants $k^E$, $k_1^\Phi$, and $k_2^\Phi$ as well as the total electric potential $\Phi$
and the radial component $E$ of the total electric field. [Note that the radial component of the electric field is the relevant one
(see Appendix~\ref{app:derivation_of_solvent_ELEs}).]
A treatment analogous to the one in Sec.~\ref{subsection:ELE} and in Appendix~\ref{app:derivation_of_solvent_ELEs}
leads to the ELEs for the four relevant profiles
\begin{align}
   \varrho_i(\bs{r})&\simeq\varrho_i^\text{b}\left[1-\beta q_i\Phi(\bs{r})\left(1+k_i^\Phi\right)\right],\;i\in\{1,2\},\label{eq:ELE_simpl1}\\
   \bar{\varrho}_0(\bs{r})&\simeq\varrho_0^\text{b},\label{eq:ELE_simpl2}\\
   f_{1,0}(\bs{r})&\simeq\frac{1}{3}\sqrt{\frac{3}{4\pi}}\beta m\left(1+k^E\right)E(\bs{r}),\label{eq:ELE_simpl3}
\end{align}
which correspond to simplified versions of Eqs.~(\ref{eq:ELE1b}), (\ref{eq:ELE_rho0}), and (\ref{eq:ELE_f10}).
Equations~(\ref{eq:ELE_simpl1})--(\ref{eq:ELE_simpl3}) together with Poisson's equation~(\ref{eq:Poisson}) lead to a linearized, modified Poisson-Boltzmann equation
\begin{align}
   \Delta\Phi(\bs{r})\simeq\frac{\displaystyle\frac{e^2I\beta}{\epsilon_0\epsilon_\text{ex}}\left(2+k_1^\Phi+k_2^\Phi\right)}{\displaystyle1+\frac{\varrho_0^\text{b}\beta m^2}{3\epsilon_0\epsilon_\text{ex}}\left(1+k^E\right)}\Phi(\bs{r})=\kappa^2\Phi(\bs{r}).
   \label{eq:mPB}
\end{align}
The requirement to recover the Debye length $1/\kappa$ in Eq.~(\ref{eq:mPB}) with
\begin{align}
   \kappa:=\sqrt{\frac{2e^2I\beta}{\epsilon_0\epsilon}},
   \label{eq:kappa}
\end{align}
and with the \textit{total} relative permittivity $\epsilon$ defines the excess relative permittivity
\begin{align}
   \epsilon_\text{ex}=\epsilon\left[1+\frac{1}{2}\left(k_1^\Phi+k_2^\Phi\right)\right]-\frac{\varrho_0^\text{b}\beta m^2}{3\epsilon_0}\left(1+k^E\right).
   \label{eq:epsex}
\end{align}
Note that in the case of vanishing particle volumes, i.e., $k_1^\Phi=k_2^\Phi=k^E=0$, and vanishing dipole moment $m=0$ the excess relative permittivity equals the
total relative permittivity $\epsilon_\text{ex}=\epsilon$.
The linearized modified Poisson-Boltzmann equation~(\ref{eq:mPB}) can be solved analytically within our geometries.

\subsection{\label{subsec:parameters}Choice of parameters}
If lengths, charges, and energies are measured in units of the Debye length $1/\kappa$ [Eq.~(\ref{eq:kappa})], the elementary
charge $e$, and the thermal energy $1/\beta=k_BT$, respectively,
the present model of a monovalent salt solution is specified by the following eleven independent, dimensionless parameters:
$\kappa r_0$, $\kappa r_1$, $\kappa r_2$, $\varrho_0^\text{b}/\kappa^3$, $I/\kappa^3$, $\kappa m/e$, $\beta u$, $\kappa r_c$,
$\kappa R$, $\sigma/(e\kappa^2)$, and $\varrho_w/\kappa^3$.
This implies that those systems are equivalent, which exhibit the same values for these dimensionless parameters.
We note that for this choice of forming dimensionless ratios the relative permittivity $\epsilon$ is not an independent parameter
but is absorbed in the expression of the Debye length [Eq.~(\ref{eq:kappa})].
The present study is focused on examining the influence of the electrode geometry.
Therefore and for illustration purposes, in the following some of these parameters are fixed to certain, realistic values, i.e.,
they are chosen such that, at best, they describe a realistic system.
\begin{table}
   \begin{ruledtabular}
   \begin{tabular}{l|ll}
      parameter                       &  case A   &   case B  \\ \hline
      $\kappa r_0$                    &  $0.1617$  &  $0.05114$  \\
      $\varrho_0^\text{b}/\kappa^3$   &  $29.36$  &  $931.8$  \\
      $I/\kappa^3$                    &  $0.05329$  &  $0.1685$  \\
      $\kappa m/e$                    &  $0.04012$  &  $0.01269$  \\
      $\beta u$                       &  $-1.500$  &  $-1.500$  \\
      $\kappa r_c$                    &  $0.5821$  &  $0.1841$  \\
      $\varrho_w/\kappa^3$            &  $9.843$  &  $311.3$  \\
   \end{tabular}
   \end{ruledtabular}
   \caption{Two sets of values for those independent, dimensionless parameters which we keep fixed for
            each case studied numerically (see Sec.~\ref{subsec:parameters}). The choice of values for the remaining dimensionless
            parameters [$\kappa r_1$, $\kappa r_2$, $\kappa R$, and $\sigma/(e\kappa^2)$] will be quoted for the corresponding numerical results.
            The choices of the values of the dimensionless parameters given in the table are guided by adopting realistic values for the corresponding
            dimensional quantities (see the main text).
            The values in case~A assume the Debye length as $1/\kappa\approx 9.600\times10^{-10}\,\text{m}$ and
            $1/\kappa\approx3.036\times10^{-9}\,\text{m}$ in case~B.
            Both cases correspond to the energy scale $\beta\approx 2.414\times10^{20}\,\text{J}^{-1}$, i.e., $T=300\,\text{K}$.
            The parameter values for case~B emerge from those of case~A by multiplying the latter by $(1/\kappa_A)/(1/\kappa_B)$ and
            $[(1/\kappa_B)/(1/\kappa_A)]^3$, respectively, and by replacing $I_A$ by $I_B=I_A/10$.
            }
   \label{tab:parameters}
\end{table}
Table~\ref{tab:parameters} provides an overview of the corresponding dimensionless parameters
for two cases~A and B. The choices of the parameter values within each case and for the two cases relative to each other are guided
by adopting realistic values for the corresponding dimensional quantities. These would, for example, refer to
an aqueous electrolyte solution at room temperature $T=300\,\text{K}$ and ambient pressure
$p\approx1013\,\text{hPa}$.
The dipole moment $m=1.85\,\text{D}\approx6.171\times10^{-30}\,\text{Cm}$ of the model solvent particles
is chosen corresponding to the literature value of $m$ for the water molecule \cite{Moore1976,CRC2017}. The relative permittivity of water in static fields
takes the value of $\epsilon=77.7003$ for our chosen temperature \cite{CRC2017}.
The equation of state for the pure solvent is derived from the functional Eq.~(\ref{eq:functional}) in the bulk
and is matched to the saturation properties of liquid water at $T=300\,\text{K}$, i.e., its saturation number density and its pressure \cite{NIST2016}.
This fixes the particle radius of the solvent to $r_0\approx 1.552\times10^{-10}\,\text{m}$.
In addition the amplitude $u$ and range $r_c$ of the attractive interaction [Eq.~(\ref{eq:Fatt})] are adjusted in order to obtain the best possible accordance
between the first peak of the structure factor of the present model and the corresponding data for water determined by X-ray scattering
(see Ref.~\cite{Franks1972} and Ref.~[783] therein). This way one obtains the values $u\approx-1.500\times k_BT$ and $r_c=3.6\times r_0$;
we recall that within our approach the interaction potentials between the substrate particles and the fluid particles are chosen to be the same square
well ones as the ones among the fluid particles.
Finally, in our model the homogeneous number density $\varrho_w$ of the particles forming the electrode enters into
the strength of the attractive interaction [Eq.~(\ref{eq:Vatt})] between the wall and the fluid particles.
Its value is estimated from the number density profile of the pure solvent in contact with an uncharged planar wall.
The choice $\varrho_w\approx1.112\times10^{28}\,\text{m}^{-3}$ ensures that the number density peak closest to the wall matches that of water
in contact with a single graphene layer \cite{Zhou2012}.
Although the value of $\varrho_w$ is expected to depend on the chosen electrode material,
the aforementioned value leads to a surprisingly good agreement also with the data corresponding to an aqueous
electrolyte solution at a charged Ag-surface (see Fig.~4a in Ref.~\cite{Toney1994}).
Case~A corresponds to an ionic strength $I=0.1\,\text{M}$, i.e., to a
Debye length $1/\kappa\approx 9.600\times10^{-10}\,\text{m}$,
whereas case~B corresponds to $I=0.01\,\text{M}$, i.e., $1/\kappa\approx3.036\times10^{-9}\,\text{m}$.
The pressure $p$ follows from the equation of state derived from the functional Eq.~(\ref{eq:functional}) in the bulk
with the equilibrium number densities $\varrho_0^\text{b}$
and $I$. The ELEs in the bulk relate the chemical potentials $\mu_i$ [Eq.~(\ref{eq:functional})]
and the thermal wave lengths $\Lambda_i$ [Eq.~(\ref{eq:Fid})], $i\in\{0,1,2\}$, with bulk quantities which have already been quoted at the beginning of
the current Subsec.~\ref{subsec:parameters}. Therefore, in this sense $\mu_i$ and $\Lambda_i$ are not independent variables.
The solvent number density $\varrho_0^\text{b}$
has to be adjusted in order to render the required value of the pressure $p$ for all examined ionic configurations. However, these variations are marginal
such that for given $\kappa$ the numerical value of $\varrho_0^\text{b}/\kappa^3$ in Tab.~\ref{tab:parameters} is valid with the precision of four significant digits.
We do not claim
that the present model is able to accurately describe liquid water because it lacks crucial properties such as hydrogen bonds and the tetrahedral shape of the
water molecules. Nevertheless, this procedure precludes one from choosing arbitrary parameter values which correspond to ``exotic'' or even unrealistic systems.
In the following this type of system, corresponding to the \textit{civ}ilized model introduced in Secs.~\ref{subsection:density_functional}--\ref{subsec:far_from_wall},
is abbreviated by ``CIV'', possibly in conjunction with additional parameter specifications or modifications
(e.g., a vanishing dipole moment, $m=0$).
In the following the remaining dimensionless parameters $r_1/r_0$, $r_2/r_0$, $\sigma/(e\kappa^2)$, and $\kappa R$
are varied and their influence on the structure of the electrolyte solution is studied.
The radii $r_1$ and $r_2$ of the ions are given in units of the radius $r_0$ of the solvent particles,
which is equivalent to providing them in units of $1/\kappa$,
and we choose either $r_1\leq r_0\leq r_2$ or $r_2\leq r_0\leq r_1$.
By choosing special values for some of the parameters, other well known models can be obtained within the described framework.
For the \textit{restricted primitive model} (RPM) one has $r_1=r_2$, $m=0$, $\beta u=0$, and $\varrho_0^\text{b}=0$,
and for the \textit{Poisson-Boltzmann} (PB) description one has $r_1=r_2=0$, $m=0$, $\beta u=0$, and $\varrho_0^\text{b}=0$.

\section{\label{section:discussion}Discussion}
\begin{figure}[t!]
   \includegraphics[width=0.44\textwidth]{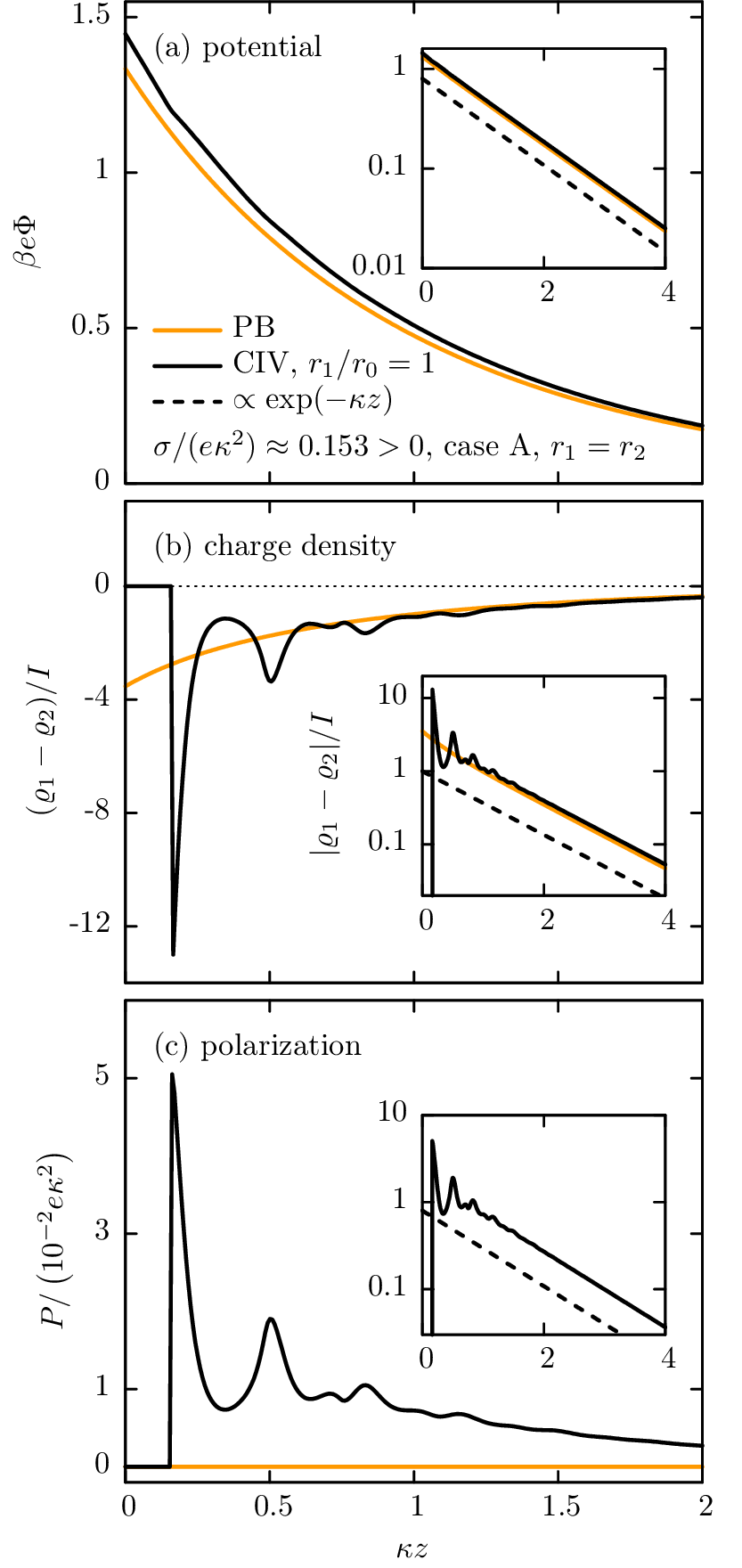}
   \caption{Reduced electrostatic potential $\beta e\Phi$ [panel (a)], reduced charge density $(\varrho_1-\varrho_2)/I$ [panel (b)],
            and component $P/(10^{-2}e\kappa^2)$ of the reduced polarization in the direction normal to the wall [Eq.~(\ref{eq:Pcomp}), panel (c)] as functions of the
            reduced distance $\kappa z$ from a planar electrode with reduced surface charge density $\sigma/(e\kappa^2)\approx0.153$.
            The CIV model and PB corresponding to case~A (see Sec.~\ref{subsec:parameters})
            are compared with each other. The insets reveal that at large distances
            from the wall the displayed profiles of both models exhibit an exponential decay on the scale of the Debye length $1/\kappa$.
            The specifications given in panel (a) apply for (b) and (c), too.}
   \label{fig:P3_010_profiles}
\end{figure}
In Fig.~\ref{fig:P3_010_profiles} various profiles relevant for the electrostatics are displayed as functions of the distance $z$ from a charged planar wall.
The CIV model, within which all particles have the same radius, and PB (see Sec.~\ref{subsec:parameters}) are compared with each other.
\begin{figure}[t!]
   \includegraphics[width=0.44\textwidth]{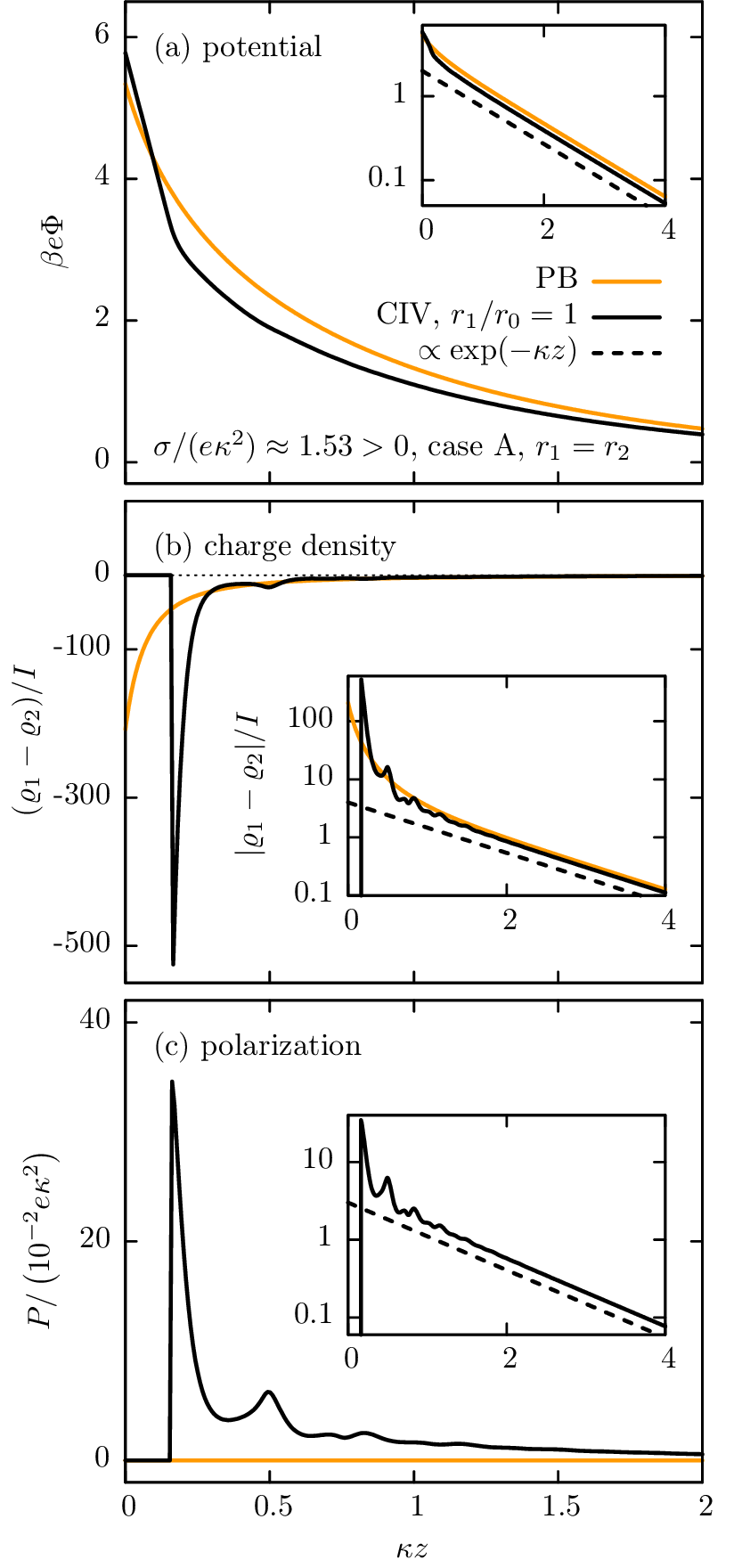}
   \caption{Same as Fig.~\ref{fig:P3_010_profiles} but for a larger value $\sigma/(e\kappa^2)\approx1.53$ of the reduced surface charge density.}
   \label{fig:P3_010_profiles_highcharge}
\end{figure}
The electrode is positively charged and consequently the electrostatic
potential $\Phi$ [Fig.~\ref{fig:P3_010_profiles}(a)] has a positive value at the wall. Qualitatively there are no significant differences in $\Phi$
between PB and CIV and even quantitatively both models lead to similar results. This is in contrast to the charge density [Fig.~\ref{fig:P3_010_profiles}(b)].
Within the microscopic CIV the centers of the fluid particles cannot get closer to the wall than their own radius.
Hence, there is a discontinuity at the distance of contact. Furthermore,
again due to the non-vanishing particle volumes, the charge density exhibits a layered structure close to the wall.%
\begin{figure}[t!]
   \includegraphics[width=0.44\textwidth]{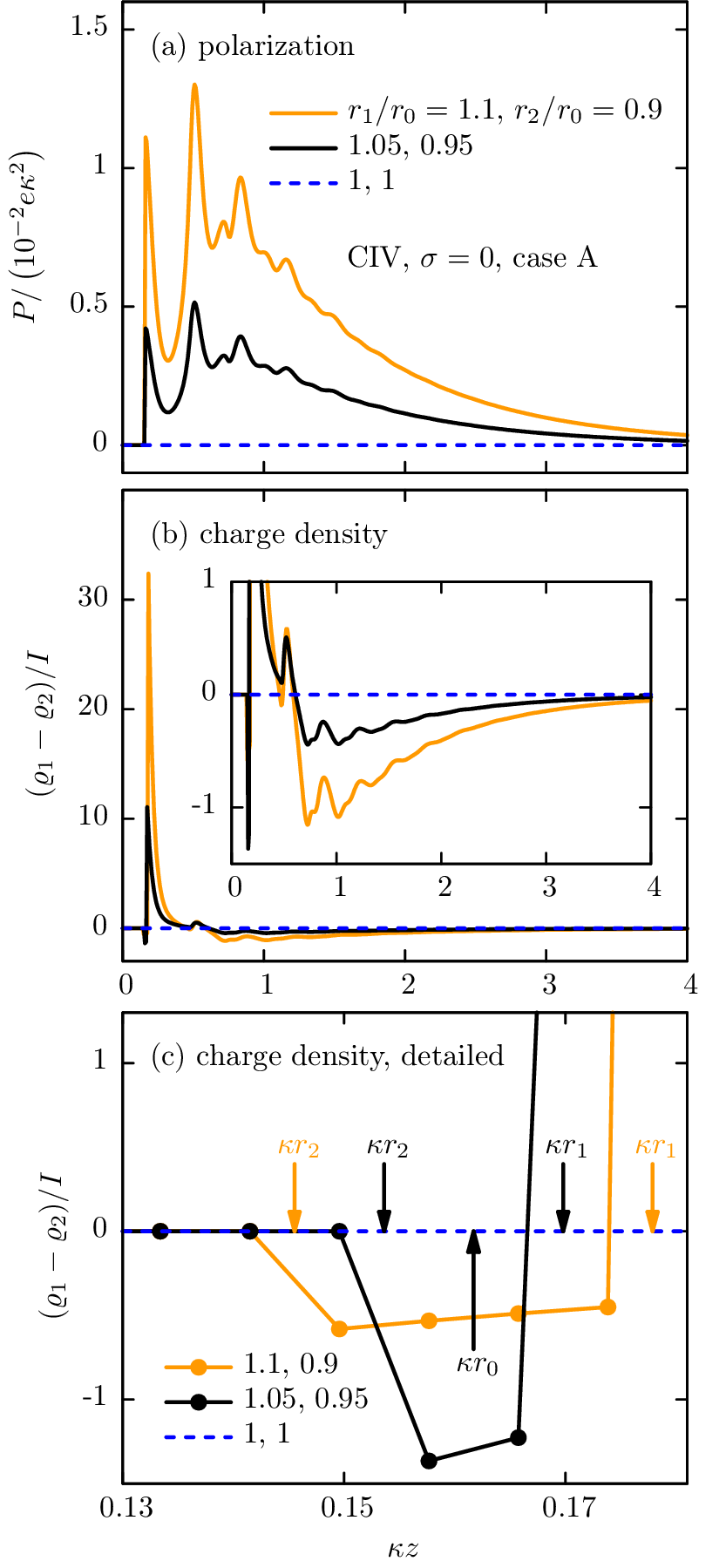}
   \caption{Component $P/(10^{-2}e\kappa^2)$ of the reduced polarization in normal direction [Eq.~(\ref{eq:Pcomp}), panel (a)]
            and reduced charge density $(\varrho_1-\varrho_2)/I$ [panels (b) and (c)] of an electrolyte solution
            as functions of the reduced distance $\kappa z$ from an uncharged planar electrode at $z=0$.
            The system is described by case~A in the CIV model (see Sec.~\ref{subsec:parameters}).
            Unequal particle radii give rise to nonzero profiles of the charge density although the wall is not charged.
            Panel (c) provides an enlarged view of the charge density close to the wall.
            There the data are plotted in the form of circles
            on the numerical grid points only and the connecting straight lines are drawn to guide the eye.
            The specifications given in panel (a) apply for (b) and (c), too.
            For further discussions of the panels and in particular of the arrows in (c) see the main text.
            \vspace{-2em}
            }
   \label{fig:P3_civ_profiles}
\end{figure}
For clarity of the presentation the solvent profile $\bar{\varrho}_0(z)$ is not shown here. However, it is noteworthy
that the high density of the solvent particles contributes considerably to the pronounced layering of the charge density. It is remarkable
that the oscillating behavior of the charge density corresponds to a rather smooth potential $\Phi$.
In contrast, in the case of PB, within which the particle volumes are
neglected, the charge density exhibits a monotonic behavior. The polarization, the component $P$ of which in the direction normal to the wall
[Eq.~(\ref{eq:Pcomp})] is shown in Fig.~\ref{fig:P3_010_profiles}(c), is identically zero in PB.
Within CIV also this profile has a layered structure close to the wall. Its positive value is in accordance with expectation because it corresponds to
dipoles, which, on average, point away from the positively charged wall. The discontinuity at contact with the wall
causes the slight kink in $\Phi$ at the same distance [see Eq.~(\ref{eq:Poisson})].
At large distances from the wall both models exhibit a monotonic exponential decay on the scale of the Debye length $1/\kappa$.

Figure~\ref{fig:P3_010_profiles_highcharge} shows results for the same models as in Fig.~\ref{fig:P3_010_profiles} but for a larger value
of the reduced surface charge density. As expected, the larger surface charge density leads to an increase of the absolute values of the shown profiles.
In addition nonlinear effects are more pronounced than in Fig.~\ref{fig:P3_010_profiles}.
This is clearly visible in the insets of panel (b): in Fig.~\ref{fig:P3_010_profiles} the PB result for the reduced charge density is almost a
straight line which is in accordance with the linearized PB equation;
in Fig.~\ref{fig:P3_010_profiles_highcharge} deviations from an exponential behavior occur.
The layering of the reduced charge density [Fig.~\ref{fig:P3_010_profiles_highcharge}(b)] and of the
component of the reduced polarization in the direction normal to the wall
[Fig.~\ref{fig:P3_010_profiles_highcharge}(c)] within the CIV model is less pronounced in the case of high surface charges. That is,
in comparison with Fig.~\ref{fig:P3_010_profiles}, the peak closest to the wall is large relative to the subsequent peaks.
For large distances from the wall, PB predicts a larger value for both the potential [Fig.~\ref{fig:P3_010_profiles_highcharge}(a)]
and the absolute value of the charge density [inset of Fig.~\ref{fig:P3_010_profiles_highcharge}(b)] than the CIV model does.
However, at contact with the electrode the order is reversed.

Figure~\ref{fig:P3_civ_profiles} shows results for the CIV model for various particle radii.
The profiles are shown as functions of the distance $z$ from an \textit{uncharged} planar wall.
If all radii are equal, i.e., $r_1/r_0=r_2/r_0=1$,
there is no electric field present and the profiles of the charge density and the polarization in Fig.~\ref{fig:P3_civ_profiles} vanish identically
due to symmetry reasons.
This changes in the case of different values of the radii. Figure~\ref{fig:P3_civ_profiles}(c)
provides an enlarged view of the charge density close to the wall.
(Due to the large zoom factor there the points of the numerical grid become visible.)
The arrows pointing downwards indicate the reduced positions of closest approach of the ions ($\kappa r_1$ and $\kappa r_2$).
The colors of the arrows and their labels correspond to the colors of the keys.
The arrow pointing upwards indicates the reduced position of closest approach of the
solvent particles ($\kappa r_0$) which is the same for all systems shown there.
The space between the electrode surface $z=0$ and the point of closest approach
($\kappa r_2$)
of the smaller ions (here negative) cannot be penetrated by any particle; in this region the charge density is identically zero.
Subsequently, in the direction away from the wall the charge density is negative because only negative ions can approach that space.
This holds up to the point beyond which the positive ions are able to penetrate that space
($\kappa r_1$); there the charge
density becomes positive. The further behavior is visible in Fig.~\ref{fig:P3_civ_profiles}(b) which shows a high but narrow positive peak.
The inset of Fig.~\ref{fig:P3_civ_profiles}(b) reveals that this positive charge is compensated by the subsequent wide region
of negative charge density such that global charge neutrality is fulfilled.
The polarization [Fig.~\ref{fig:P3_civ_profiles}(a)] has a discontinuity at the position
$z=r_0$, i.e., at the point of closest approach of the solvent particles.
In Fig.~\ref{fig:P3_civ_profiles}(c), the arrow pointing upwards indicates this position $\kappa r_0$.
Because the value of the radius of the
solvent particles is chosen to be in between the values of the radii of the ions, the discontinuity of the polarization $(\kappa r_0)$ is located
to the right of the point of closest approach of the negative ions $(\kappa r_2)$ and to the left of the point of closest approach of the positive ions
$(\kappa r_1)$.

In the following we discuss the properties of an electrolyte solution in contact with an electrode in terms of the differential capacitance
\cite{Schmickler2010}
\begin{align}
   C:=\frac{\partial\sigma}{\partial\Phi(\bs{r})|_{\mathcal{A}}}
   \label{eq:C}
\end{align}
which is the change of the surface charge density $\sigma$ upon varying the (constant) potential at the wall
$\Phi(\bs{r})|_{\mathcal{A}}$ taken relative to its bulk value.
Within the present study the ELEs in Sec.~\ref{subsection:ELE} are solved for various values of the surface charge density $\sigma$.
Together with the solutions of the ELEs the electric potential $\Phi$ [Eq.~(\ref{eq:Phi})] is known.
The relation between the potential at the wall and $\sigma$ is used in order to determine $C$ [Eq.~(\ref{eq:C})] numerically.
The differential capacitance is experimentally accessible, e.g., via cyclic voltammetry, chronoamperometry, and impedance spectroscopy \cite{Butt2003}.
It contains integrated properties of the structure of
the electrolyte solution (see, e.g., Figs.~\ref{fig:P3_010_profiles}--\ref{fig:P3_civ_profiles}).
This facilitates the comparison with other
models and the analysis of the influence of the parameters
$r_1/r_0$, $r_2/r_0$, $\sigma/(e\kappa^2)$, $I/\kappa^3$, and $\kappa R$.

\begin{figure}[t!]
   \includegraphics[width=0.44\textwidth]{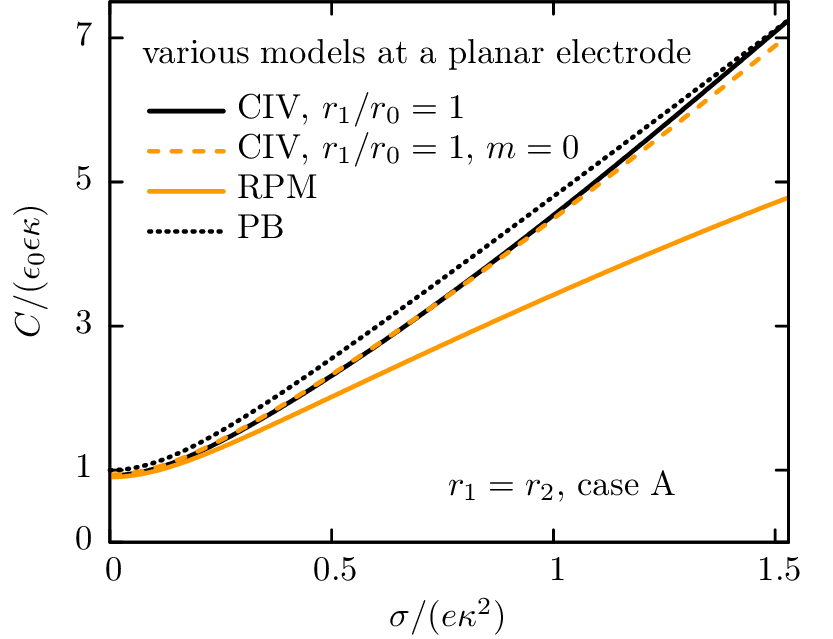}
   \caption{Reduced differential capacitance $C/(\epsilon_0\epsilon\kappa)$ as a function of the reduced surface charge density
            $\sigma/(e\kappa^2)$ of a planar electrode.
            The electrolyte solution corresponding to case~A is described as indicated by various models in which the ionic radii are chosen to be equal
            ($r_1=r_2$, see Sec.~\ref{subsec:parameters}).}
   \label{fig:P3_cap_pla1}
\end{figure}
Figure~\ref{fig:P3_cap_pla1} depicts results of various models in which the ionic radii are chosen to be equal
($r_1=r_2$, see Sec.~\ref{subsec:parameters}).
The electrolyte solution is in contact with a planar electrode the rescaled surface charge density $\sigma$ of which is the horizontal axis.
Here and in the following, the differential capacitance is plotted in units of the double-layer capacitance $\epsilon_0\epsilon\kappa$ which
facilitates comparison with Gouy-Chapman results (see also Ref.~\cite{Reindl2016}).
Within the range shown, all displayed curves exhibit the same characteristics as the PB result: for small $\sigma$ the differential capacitance
attains a constant with zero slope.
Upon increasing $\sigma$ also the capacitance increases; the main differences between the models are borne out within this range.
The two indicated CIV results differ with respect to the strength of the dipole moment $m$: in one case (black solid line) the latter is chosen according
to Sec.~\ref{subsec:parameters} and in the other case (orange dashed line) it is set to zero.
The two curves almost coincide and only for large values of $\sigma$ a small deviation is visible.
Hence, for these two systems the influence of the dipole moment is relatively
weak. This finding is in accordance with previous studies of the differential capacitance \cite{Carnie1980,Warshavsky2016}
as well as of wetting phenomena \cite{Oleksy2010} for which
the explicit dipole description turned out to have a relatively small effect.
The agreement between the simple PB and the comparatively complex CIV models throughout the studied range is remarkable,
in particular when taking into account
that RPM, endowed with an intermediate degree of complexity, clearly shows deviations from the otherwise common trend.
An explanation for this observation could be that within CIV and PB all particles of 
the electrolyte solution are described consistently on the same footing whereas within RPM they are not:
within PB all particles are pointlike and within the displayed cases of CIV
the particles are treated as hard spheres of \textit{equal} radii. In contrast, within RPM the solvent is a structureless continuum and
the ions are described as hard spheres of finite size.

\begin{figure}[t!]
   \includegraphics[width=0.44\textwidth]{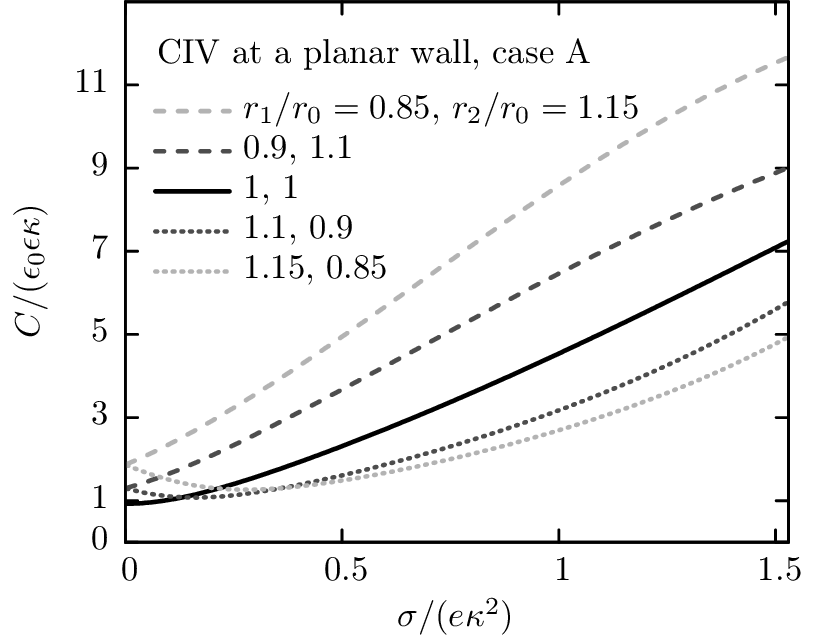}
   \caption{Reduced differential capacitance $C/(\epsilon_0\epsilon\kappa)$ as a function of the reduced surface charge density
            $\sigma/(e\kappa^2)$ of a planar electrode. The electrolyte solution is
            described within the CIV model in case~A (see Sec.~\ref{subsec:parameters}).
            Curves with the same shade of gray correspond to the same \textit{set} of ion size ratios $\{r_1/r_0,r_2/r_0\}$ where the cations are the larger ions
            $(r_1/r_0>r_2/r_0)$ for dotted curves whereas the anions are the larger ions $(r_1/r_0<r_2/r_0)$ for dashed curves.
            The PB result turns out to be close to the solid black curve (see Fig.~\ref{fig:P3_cap_pla1}).}
   \label{fig:P3_cap_pla2}
\end{figure}
In Fig.~\ref{fig:P3_cap_pla2} the influence of various choices for the ionic radii is examined within CIV: the solid curve corresponds to the case in which all radii
are equal whereas they are unequal for the other curves. In the limit of an uncharged ($\sigma\rightarrow0$) electrode the capacitance increases with
increasing difference between the ionic radii.
Curves with the same shade of gray correspond to the same \textit{set} of ion size ratios $\{r_1/r_0,r_2/r_0\}$ where the cations are the larger ions
$(r_1/r_0>r_2/r_0)$ for dotted curves whereas the anions are the larger ions $(r_1/r_0<r_2/r_0)$ for dashed curves.
As expected, curves of the same shade concur at the vertical axis:
For $\sigma\rightarrow0$ swapping the ion radii is equivalent to flipping the sign of all charges and the latter does not change the differential capacitance
$C$ [see Eq.~(\ref{eq:C})].
However, for a charged electrode ($\sigma>0$) this equivalence does not hold
and for the differential capacitance two branches occur.
If the positive ions are the smaller (larger) particles, the capacitance curve exhibits (non-)monotonic behavior
within the investigated interval of $\sigma$.
Compared to the models in Fig.~\ref{fig:P3_cap_pla1}, where the ionic radii are chosen to be equal, the decrease of $C$
for small values of $\sigma$ is a new feature in Fig.~\ref{fig:P3_cap_pla2} for $r_1/r_0>1$ and $r_2/r_0<1$.
On the other hand, for $r_1/r_0<1$ and $r_2/r_0>1$ the capacitance increases rapidly thus
leading to values of $C$ larger than those for the cases shown in Fig.~\ref{fig:P3_cap_pla1}.
Due to the symmetries of the present model the differential capacitance fulfills the relation $C(r_1,r_2,\sigma)=C(r_2,r_1,-\sigma)$,
where the first, second, and third argument are the cation radius, the anion radius, and the surface charge density, respectively.
That is, if the capacitance is known for a particular system, the same capacitance is obtained for a system in which the values of the ionic
radii are swapped and the surface charge density is taken to be opposite. This explains why curves with the same shade of gray
meet at $\sigma=0$ in Fig.~\ref{fig:P3_cap_pla2}. Moreover the above relation enables one to extend the curves in Fig.~\ref{fig:P3_cap_pla2} to
negative values of $\sigma$. For unequal values of the ionic radii the resulting curve has a minimum at a certain nonzero value of $\sigma$
and the curve is not symmetric with respect to this minimum. The shape of such a curve is in better qualitative agreement with experimental findings
than the curve for $r_1=r_2$ which is symmetric around the minimum at $\sigma=0$ (see, e.g., Ref.~\cite{Valette1981}). 

\begin{figure}[t!]
   \includegraphics[width=0.44\textwidth]{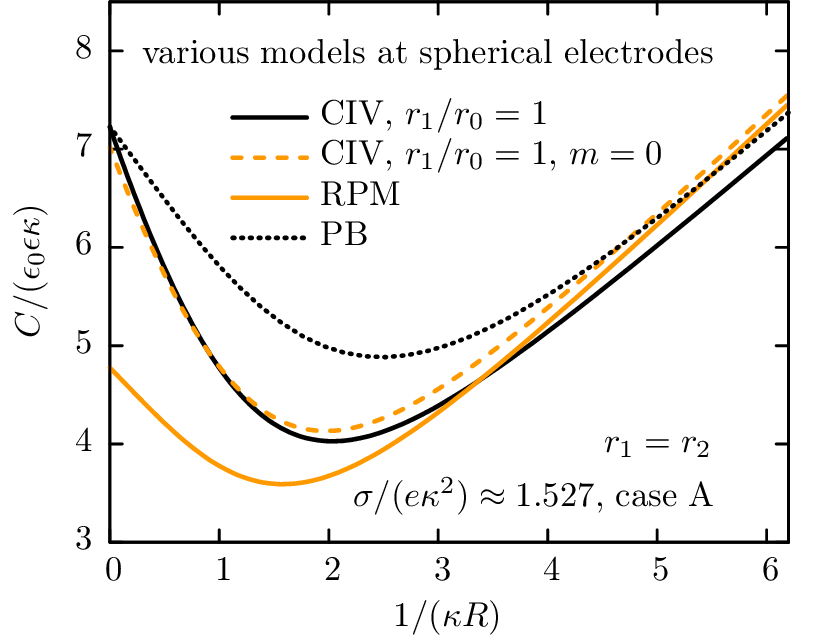}
   \caption{Reduced differential capacitance $C/(\epsilon_0\epsilon\kappa)$ as a function of the
            dimensionless curvature $1/(\kappa R)$ of spherical electrodes.
            The electrolyte solution corresponding to case~A (see Sec.~\ref{subsec:parameters})
            is described by the same models as the ones used in Fig.~\ref{fig:P3_cap_pla1}.
            The value of the reduced surface charge density is $\sigma/(e\kappa^2)\approx1.527$.
            The planar limit [$1/(\kappa R)=0$] corresponds to that planar electrode in Fig.~\ref{fig:P3_cap_pla1}
            with the highest surface charge density.}
   \label{fig:P3_cap_sph1}
\end{figure}
So far the focus has been on planar electrodes (see Figs.~\ref{fig:P3_010_profiles}--\ref{fig:P3_cap_pla2}). In Fig.~\ref{fig:P3_cap_sph1} the same models as in Fig.~\ref{fig:P3_cap_pla1}
are used in order to investigate electrolyte solutions in contact with spherical electrodes of various radii $R$. The surface charge density of $\sigma/(e\kappa^2)\approx1.527$
is chosen such that the planar limit, i.e., for zero curvature $1/(\kappa R)=0$, corresponds to that planar electrode in Fig.~\ref{fig:P3_cap_pla1}
with the largest surface charge density.
In this planar limit PB and CIV (dotted and solid black line in Fig.~\ref{fig:P3_cap_pla1}, respectively)
yield almost the same value for the differential capacitance. However, in Fig.~\ref{fig:P3_cap_sph1} differences 
between the two models appear for nonzero curvatures, i.e., finite electrode radii: PB predicts larger values for the capacitance than CIV.
Hence, compared with the situation at planar electrodes, where PB is a surprisingly accurate approximation for CIV with equal particle radii,
at curved electrodes larger deviations occur.
It is likely that these differences originate from the hard-sphere character of the particles within CIV.
As within PB, the charged electrode interacts with the charges of the ions and denies them access to a certain $R$-dependent volume.
However, in the case of CIV in addition
the layering of the particles is influenced by varying the radius of the electrode. A mechanism of such kind is not present within PB.
This might explain, why between CIV and PB there are differences in the curvature dependences 
and why the importance of microscopic details hinges on the geometry of the electrode.
Again (as in Fig.~\ref{fig:P3_cap_pla1}) the two CIV results shown in Fig.~\ref{fig:P3_cap_sph1} are close to each other
indicating that within CIV the dipole moment has no significant effect on the capacitance.
Since already at a planar wall RPM exhibits clear deviations from the other models (see Fig.~\ref{fig:P3_cap_pla1}),
it does not come as a surprise, that the curve predicted by it deviates considerably also at spherical electrodes
(see Fig.~\ref{fig:P3_cap_sph1}). For small wall radii the various models
seem to attain a linear dependence on curvature and, compared with intermediate values of the curvature, these
lines are relatively close to each other.

\begin{figure}[t!]
   \includegraphics[width=0.44\textwidth]{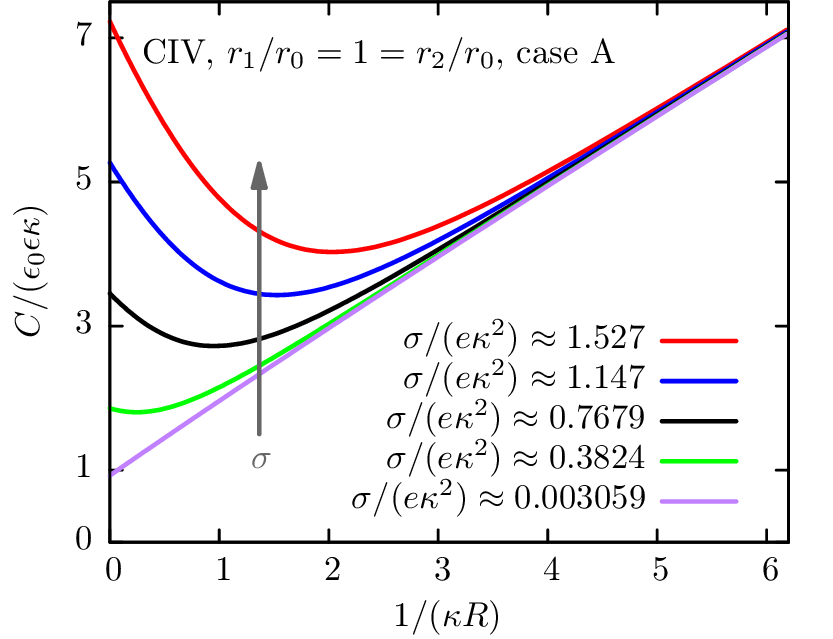}
   \caption{Reduced differential capacitance $C/(\epsilon_0\epsilon\kappa)$ as a function of the dimensionless curvature
            $1/(\kappa R)$ of spherical electrodes. The displayed data correspond to an electrolyte
            solution described by case~A in the CIV model (see Sec.~\ref{subsec:parameters}) with equal particle radii $r_1/r_0=1=r_2/r_0$.
            Each curve corresponds to a certain value of the surface charge
            density $\sigma$. The vertical arrow indicates the direction of increasing $\sigma$.
            }
   \label{fig:P3_010}
\end{figure}
\begin{figure}[t!]
   \includegraphics[width=0.44\textwidth]{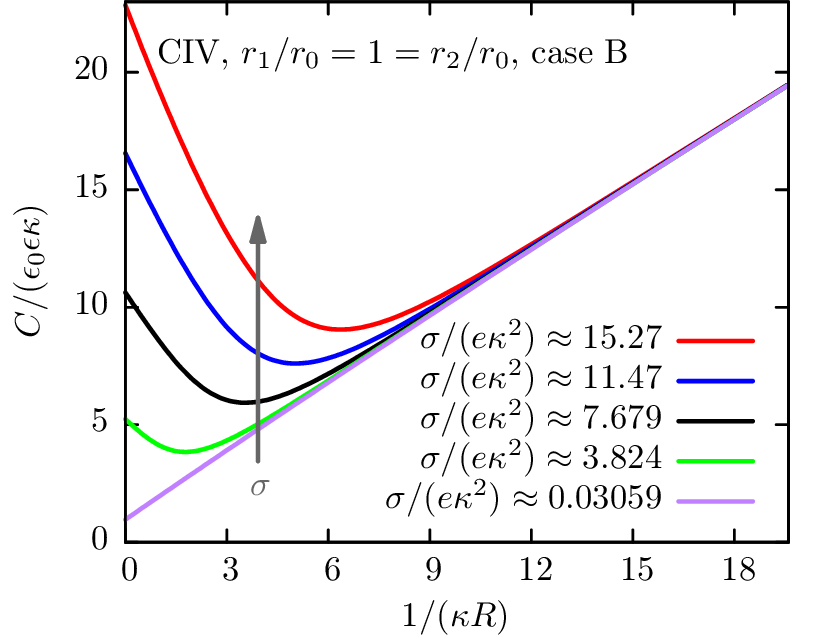}
   \caption{Same as Fig.~\ref{fig:P3_010} for case~B (see Sec.~\ref{subsec:parameters}).}
   \label{fig:P3_020}
\end{figure}
\begin{figure}[t!]
   \includegraphics[width=0.44\textwidth]{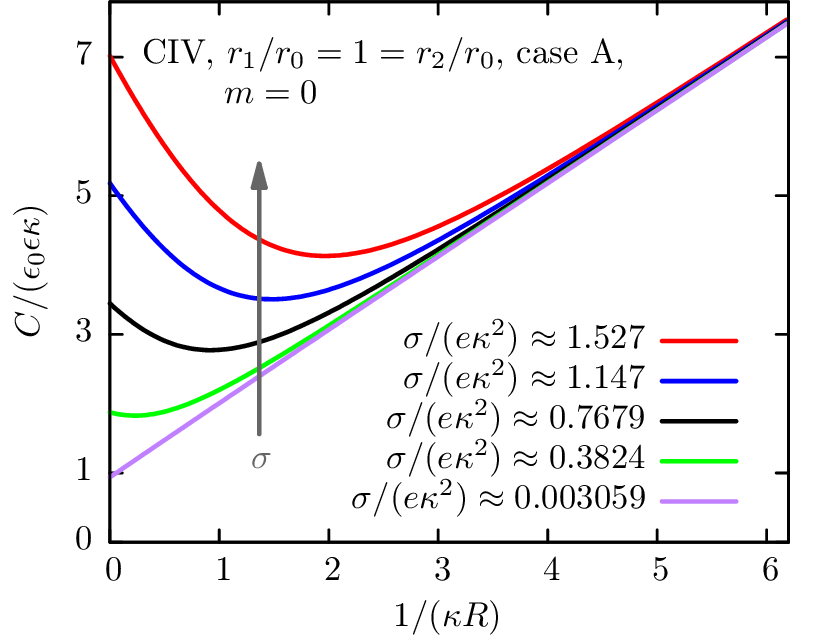}
   \caption{Same as Fig.~\ref{fig:P3_010} for $m=0$.}
   \label{fig:P3_030}
\end{figure}
\begin{figure}[t!]
   \includegraphics[width=0.44\textwidth]{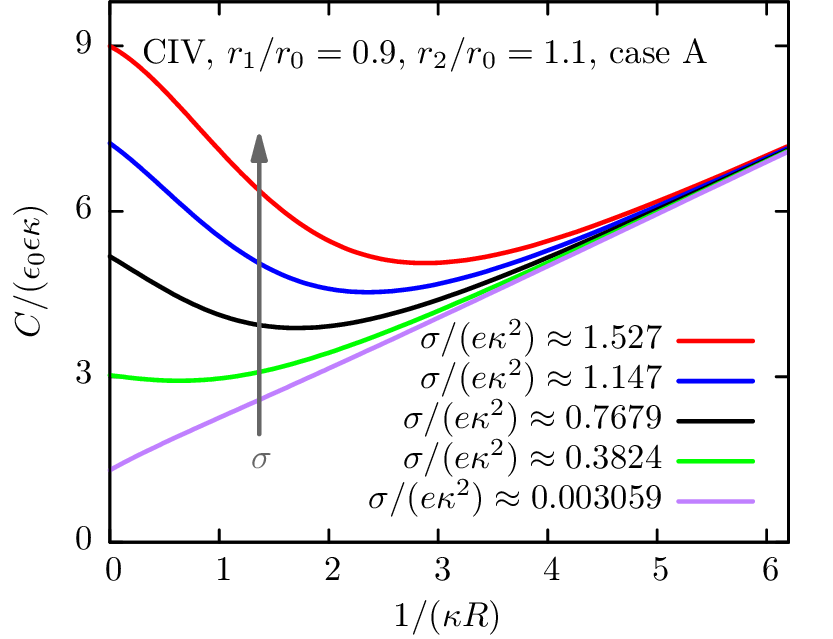}
   \caption{Same as Fig.~\ref{fig:P3_010} for $r_1/r_0=0.9$ and $r_2/r_0=1.1$.}
   \label{fig:P3_104}
\end{figure}
\begin{figure}[t!]
   \includegraphics[width=0.44\textwidth]{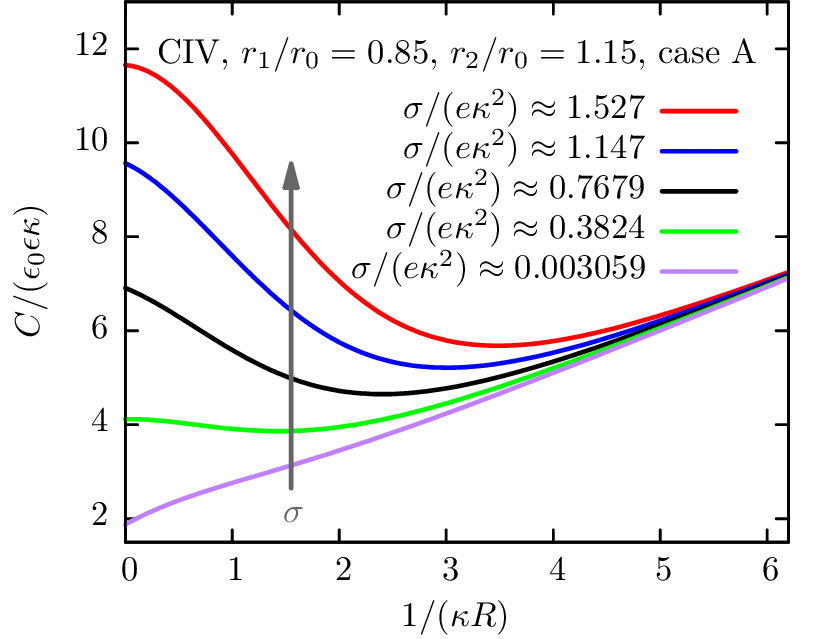}
   \caption{Same as Fig.~\ref{fig:P3_010} for $r_1/r_0=0.85$ and $r_2/r_0=1.15$.}
   \label{fig:P3_101}
\end{figure}
\begin{figure}[t!]
   \includegraphics[width=0.44\textwidth]{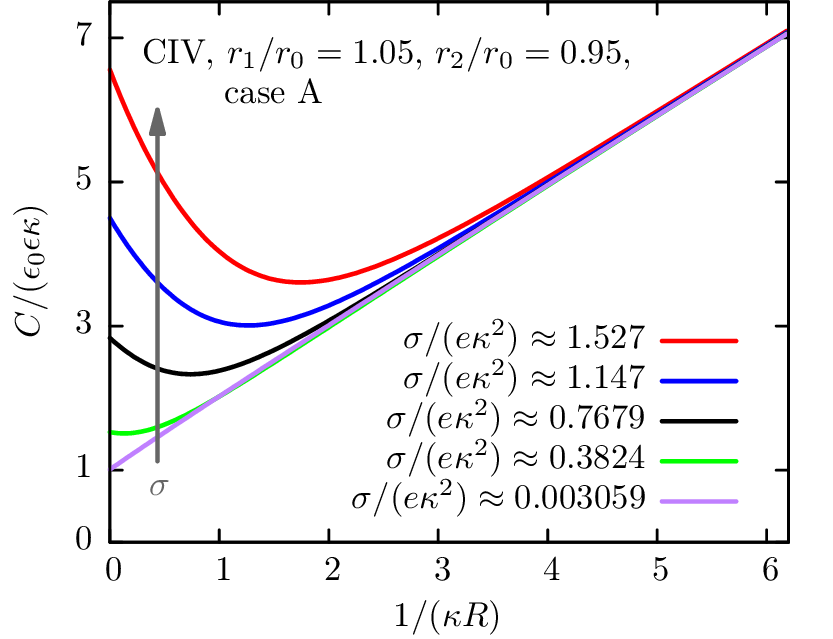}
   \caption{Same as Fig.~\ref{fig:P3_010} for $r_1/r_0=1.05$ and $r_2/r_0=0.95$.}
   \label{fig:P3_102}
\end{figure}
\begin{figure}[t!]
   \includegraphics[width=0.44\textwidth]{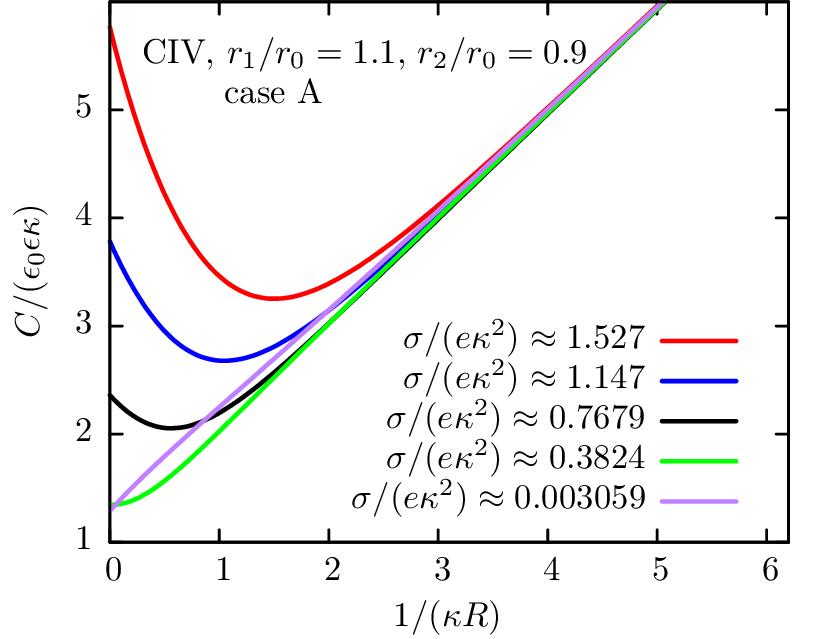}
   \caption{Same as Fig.~\ref{fig:P3_010} for $r_1/r_0=1.1$ and $r_2/r_0=0.9$.
            The only concave curve corresponds to the bottom entry of the key;
            the order of the vertical intercepts of the convex curves corresponds
            to the order of the remaining keys.
           }
   \label{fig:P3_105}
\end{figure}
\begin{figure}[t!]
   \includegraphics[width=0.44\textwidth]{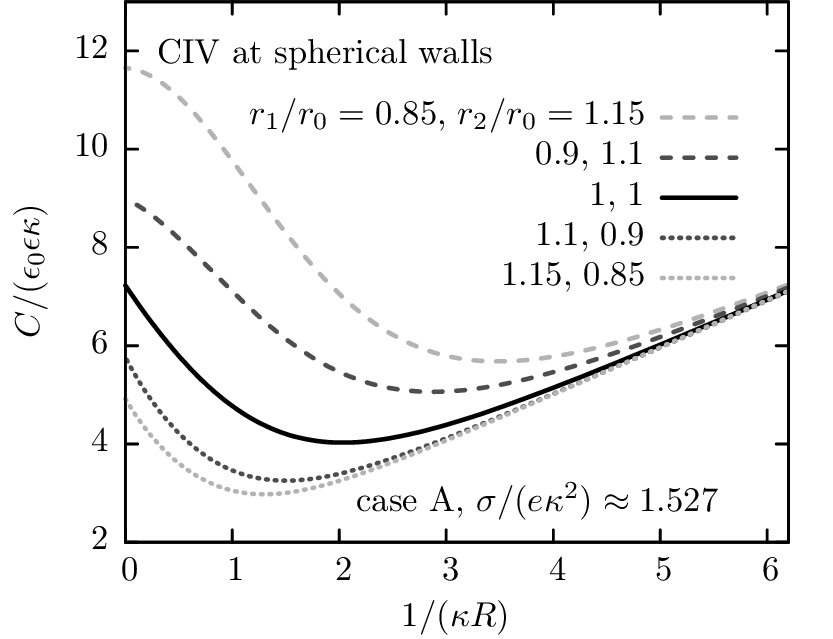}
   \caption{Reduced differential capacitance $C/(\epsilon_0\epsilon\kappa)$ as a function of the dimensionless curvature $1/(\kappa R)$ of spherical electrodes
            with rescaled surface charge density $\sigma/(e\kappa^2)\approx1.527$.
            The electrolyte solution corresponding to case~A is described within the CIV model (see Sec.~\ref{subsec:parameters}).
            Curves with the same shade of gray correspond to the same \textit{set} of ion size ratios $\{r_1/r_0,r_2/r_0\}$ where the cations are the larger ions
            $(r_1/r_0>r_2/r_0)$ for dotted curves and the anions are the larger ions $(r_1/r_0<r_2/r_0)$ for dashed curves.}
   \label{fig:P3_cap_sph2}
\end{figure}

Figures~\ref{fig:P3_010}--\ref{fig:P3_105} display the differential capacitance $C$ of spherical electrodes
as a function of their curvature $1/(\kappa R)$.
The corresponding data are obtained within CIV (see Sec.~\ref{subsec:parameters}) and each curve corresponds
to a fixed value of the surface charge density $\sigma$.
For $1/(\kappa R)=0$ the capacitance values
reduce to the corresponding ones for a planar wall. For the largest curvatures considered, the radius of the electrode approximately
equals the radii of the fluid particles.
It is remarkable that all systems studied exhibit a common behavior for \textit{large} curvatures:
irrespective of the value of $\sigma$ all curves converge to the graph
corresponding to $\sigma\rightarrow0$. However, the dependence on $\sigma$ becomes non-trivial for small curvatures. 
A similar general behavior for \textit{large} curvatures has been observed in part I \cite{Reindl2016} of our study,
where the PB model is discussed in detail.
There it is possible to show analytically, that for sufficiently small radii (i.e., large curvatures) of the electrode the linearized version of
the PB equation is a reliable description. Within the linearized PB theory the electrode potential is proportional to the surface charge density 
and hence the differential capacitance is independent of $\sigma$.
This explains within PB, why at large curvatures the dependence of the capacitance $C$ on $\sigma$ disappears.
It is not possible to analytically analyze the CIV model as detailed as PB. However, the data from the CIV model in the present study reveal
the same behavior as in the PB model, i.e., the dependence of $C$ on $\sigma$ weakens for large curvatures.
Furthermore, Fig.~\ref{fig:P3_cap_sph2} demonstrates that the capacitance $C$
at large curvatures is also independent of the radii of the particles. Hence the capacitance exhibits a general
behavior for \textit{large} curvatures which is independent of $\sigma$ and of the particle radii.

Figures~\ref{fig:P3_010}--\ref{fig:P3_030} are related because in the cases studied there the radii of all particles are chosen to be equal.
Compared with Fig.~\ref{fig:P3_010}, in Fig.~\ref{fig:P3_020} the ionic strength is reduced, and in Fig.~\ref{fig:P3_030} the dipole moment is set to zero.
Qualitatively these three systems exhibit
similar curves and their shapes resemble the results obtained within the pure PB approach in part I \cite{Reindl2016} of this study.
Quantitatively, however, the distinct models render deviations which are visible
most clearly in Fig.~\ref{fig:P3_cap_sph1}, where various approaches are compared for one fixed value of the surface charge density.
Again, the results are only weakly affected by the strength of the dipole moment:
the graphs in Fig.~\ref{fig:P3_010} (with the dipole moment chosen as in Sec.~\ref{subsec:parameters})
and Fig.~\ref{fig:P3_030} (no dipole moment) differ only slightly. Also in the case of planar walls (see Fig.~\ref{fig:P3_cap_pla1}
and Refs.~\cite{Carnie1980,Oleksy2010,Warshavsky2016})
the explicit dipole description turned out to have only a small effect.
It would be interesting to counter-check this finding with alternative approaches such as computer simulations.
Thereby it might be possible to clarify, whether the aforementioned small differences in the capacitance
originate from an insufficient description of the dipoles or whether already 
simpler models are capable to capture sufficiently accurately the relevant structure of an electrolyte solution.
In this case it might be justified to skip
the comparatively sophisticated description of the dipoles. Simulations for models, which take dipoles explicitly into account,
have already been carried out.%
In Ref.~\cite{Boda1998} results for mixtures of hard spherical ions and dipoles
in contact with charged walls are presented in terms of spatially varying profiles.
Reference~\cite{Pegado2016} summarizes several
simulation studies concerning the effective interaction between two charged surfaces separated by a solution described by ions dissolved in a Stockmayer fluid
which is a Lennard-Jones fluid with an embedded point-dipole.
However, computer simulations of ion-dipole mixtures are regarded to be technically difficult \cite{Boda1998}.
Possibly, this is the reason why, to our knowledge, numerical capacitance data derived from models with and without explicit dipole description
had not yet been compared with each other, as it is done, e.g., in Fig.~\ref{fig:P3_cap_pla1}.

Unequal particle radii can give rise to qualitatively different behaviors which may be discussed according to the sizes of the ionic species.
In the case that the positive ions are the smaller particles (Figs.~\ref{fig:P3_104} and \ref{fig:P3_101})
the capacitance in the planar limit $1/(\kappa R)\rightarrow0$ increases with increasing difference in the particle radii.
Moreover the graphs become more concave (from below) for small curvatures and in particular for large surface charge densities.
In the case that the positive ions are the largest particles (Figs.~\ref{fig:P3_102} and \ref{fig:P3_105})
the capacitance in the planar limit shows a more complex
behavior (see also Fig.~\ref{fig:P3_cap_pla2}): for $\sigma\rightarrow0$ the capacitance increases with increasing difference in the particle radii.
However, for intermediate and large values of $\sigma$, the capacitance
decreases upon increasing the particle size difference. As a consequence, some curves approach the graph for $\sigma\rightarrow0$ from \textit{below}
(see, e.g., the (green) graph for $\sigma/(e\kappa^2)\approx0.3824$ in Fig.~\ref{fig:P3_105}),
whereas in the most other cases the convergence is from above
(see Figs.~\ref{fig:P3_010}--\ref{fig:P3_101}).

Already in the case of planar electrodes it has become apparent that a variation of the particle radii has a relatively strong effect on the shape of capacitance data
(see Figs.~\ref{fig:P3_cap_pla1} and \ref{fig:P3_cap_pla2}). This finding is confirmed for the case of spherical electrodes when comparing the data
corresponding to unequal particle radii (Figs.~\ref{fig:P3_104}--\ref{fig:P3_105})
with the data corresponding to equal particle radii (Figs.~\ref{fig:P3_010}--\ref{fig:P3_030}), upon varying the ionic strength $I$ or the dipole moment $m$.
The difference in particle radii has a strong influence on
the charge distribution because the smallest species can approach the wall closest. For a planar wall this behavior is captured in Fig.~\ref{fig:P3_civ_profiles}.
For spherical electrodes the interplay of these steric effects with electrostatic interactions is influenced additionally by the radius $R$ of the electrode which increases the
complexity and gives rise to the various shapes of the presented capacitance data. Figure~\ref{fig:P3_cap_sph2} facilitates the comparison of distinct data sets.
The solid curve corresponds to the case in which all radii are equal whereas the radii are unequal for the cases corresponding to
the other curves. For large curvatures the curves approach
each other and exhibit a common behavior independent of the chosen particle radii.
In view of the limiting behavior at large curvatures, as shown in Figs.~\ref{fig:P3_010}--\ref{fig:P3_105},
the common behavior at large curvatures is also independent of the surface charge density.
At this stage it is already known that the simple PB model is a rather good approximation for CIV with equal particle radii in the 
limit of small electrode radii (see Fig.~\ref{fig:P3_cap_sph1}).
Furthermore, in part I \cite{Reindl2016} of this study it is shown that in the limit of large curvatures these results 
are in accordance with the \textit{linearized} PB description. Combined with the insight obtained from Fig.~\ref{fig:P3_cap_sph2} it seems that for $1/(\kappa R)\gg1$ the
linearized PB model might be an adequate approximation for all systems displayed in Fig.~\ref{fig:P3_cap_sph2}.
This finding might be interesting for describing small electrodes or highly curved parts of electrodes.

\section{\label{section:summary}Summary}
We have analyzed the electric double layer (EDL) of an electrolyte solution in contact with electrodes of planar or spherical shape.
Inspired by the study of Oleksy and Hansen \cite{Oleksy2010} the electrolyte solution is described in terms of density functional theory (DFT)
based on the functional given in Eq.~(\ref{eq:functional}). This approach, which is a certain version of the
so-called \textit{civ}ilized model (CIV, see Sec.~\ref{subsec:parameters}), takes into account all particle species on equal footing.
All particles are modelled as hard spheres with non-vanishing
volumes, embedded charges (in the cases of the monovalent anions or cations) or point-dipoles (in the case of the solvent molecules),
and with an attractive interaction amongst all particles
which enables one to discuss an electrolyte solution in the liquid state under realistic ambient conditions.
This microscopic model is a possible extension of the mesoscopic
Poisson-Boltzmann (PB) approach, which was used in part I of our study \cite{Reindl2016} in order to discuss EDLs at curved electrodes.
Close to the wall the microscopic description gives rise to a layering behavior of the charge density and of the polarization
(see Figs.~\ref{fig:P3_010_profiles}--\ref{fig:P3_civ_profiles}) whereas the PB approach renders monotonic profiles only. 
As in part I \cite{Reindl2016} the structural features of the EDL enter into the differential capacitance $C$ [Eq.~(\ref{eq:C})] which facilitates
the comparison of various models with each other or to evaluate the 
influence of various system parameters such as particle radii, dipole moment of the solvent molecules, ionic strength, surface charge density, and electrode radius.
At the planar wall and for equal radii of all particles, PB and CIV lead to similar values for the capacitance (see Fig.~\ref{fig:P3_cap_pla1}).
Since compared with CIV (see Sec.~\ref{subsec:parameters}) PB neglects many microscopic details, this finding is not obvious.
Against this background, in its turn it is remarkable, that in the case of spherical electrodes of finite radii $R$ the agreement between the
predictions of the two models
deteriorates (see Fig.~\ref{fig:P3_cap_sph1}), i.e., the relevance of microscopic details, captured by the CIV model,
depends on the geometry of the electrode.
The restricted primitive model (RPM), in which the particles are not treated on equal footing,
clearly exhibits a different trend in comparison with the other models (see Figs.~\ref{fig:P3_cap_pla1} and \ref{fig:P3_cap_sph1}).
In the case of spherical electrodes the capacitance data obtained within CIV for equal particle radii are
qualitatively similar to the PB results of part I \cite{Reindl2016} of this study
(see Figs.~\ref{fig:P3_010}--\ref{fig:P3_030}). Nevertheless there are quantitative differences (see Fig.~\ref{fig:P3_cap_sph1}).
Considering the dipoles explicitly has no large effect
(compare the two curves labelled with CIV in Figs.~\ref{fig:P3_cap_pla1} and \ref{fig:P3_cap_sph1}
or compare Fig.~\ref{fig:P3_010} with Fig.~\ref{fig:P3_030}).
Qualitative and relatively large quantitative differences occur if the particle radii are unequal.
This is the case both for planar electrodes [see Fig.~\ref{fig:P3_cap_pla2} where the PB result turns out to be close to the solid black curve
(see Fig.~\ref{fig:P3_cap_pla1})]
and for spherical electrodes (compare Fig.~\ref{fig:P3_010} with Figs.~\ref{fig:P3_104}--\ref{fig:P3_105}).
However, the differences are borne out only for small and intermediate curvatures $1/(\kappa R)$. For large curvatures the capacitance curves
of all considered cases exhibit a common behavior and converge to the limiting graph valid
for small surface charge densities $\sigma\rightarrow0$, i.e., in this limit 
the behavior becomes independent of $\sigma$ (see Figs.~\ref{fig:P3_010}--\ref{fig:P3_105}).
Moreover, this behavior becomes also independent of the choice of the particle radii (see Fig.~\ref{fig:P3_cap_sph2}).
For $1/(\kappa R)\gg1$ the simple linearized PB model appears to be an adequate approximation of the relatively complex CIV.
We conclude that the geometry of the electrode determines the relevance of microscopic details.
Apart from the limit of small electrode radii, for which a general
behavior is observed, PB provides acceptable estimates in the case of equal particle sizes and large electrode radii.


\appendix

\begin{widetext}

\section{\label{app:derivation_of_solvent_ELEs}Derivation of the ELEs for the solvent in the form of Eqs.~(\ref{eq:ELE_rho0}) and (\ref{eq:ELE_f10})}
Equations~(\ref{eq:ELE_rho0}) and (\ref{eq:ELE_f10}) follow from the ELE~(\ref{eq:ELE1a}) which contains all information needed about the solvent.
It can be written as
\begin{align}
   \begin{aligned}
      \varrho_0(\bs{r},\bs{\omega})&=\zeta(\bs{r})\exp\left\{\beta m\bs{\omega}\cdot\left[\bs{E}(\bs{r})-\bs{E}^\text{aux}(\bs{r})\right]\right\},\\
                      \zeta(\bs{r})&:=\frac{\varrho_0^\text{b}}{4\pi}\exp\left[-\beta V_0(\bs{r})+c_0^\text{hs}(\bs{r})-c_0^\text{hs,b}+c_0^\text{att}(\bs{r})-c_0^\text{att,b}\right].
   \end{aligned}
   \label{eq:app1_0}
\end{align}
Due to the dependence of Eq.~(\ref{eq:app1_0}) on both the position $\bs{r}$ and the orientation $\bs{\omega}$, in general the equation
has to be solved in a high-dimensional space. In order to reduce the dimensionality of the problem
we focus only on the relevant information contained in Eq.~(\ref{eq:app1_0}).
With Eq.~(\ref{eq:app1_0}) and the definition of the orientationally integrated number density
$\bar{\varrho}_0(\bs{r})$ of the solvent [Eq.~(\ref{eq:rho0_all_dipoles})] one has
\begin{align}
   \bar{\varrho}_0(\bs{r})=\zeta(\bs{r})\int \ud^2\omega\,\exp\left\{\beta m\bs{\omega}\cdot\left[\bs{E}(\bs{r})-\bs{E}^\text{aux}(\bs{r})\right]\right\}.
   \label{eq:app1_1}
\end{align}
In order to carry out the angular integration in Eq.~(\ref{eq:app1_1}) the orientation vector $\bs{\omega}$ [Eq.~(\ref{eq:omega})]
is represented in a coordinate system the polar axis of which points into the radial direction away from the wall, i.e., the polar axis is parallel to the electric fields
$\bs{E}(\bs{r})$ and $\bs{E}^\text{aux}(\bs{r})$.
Therefore the scalar product in Eq.~(\ref{eq:app1_1}) reduces to
\begin{align}
   \begin{aligned}
      \beta m\bs{\omega}\cdot\left[\bs{E}(\bs{r})-\bs{E}^\text{aux}(\bs{r})\right]=\beta m\left[E(\bs{r})-E^\text{aux}(\bs{r})\right]\cos(\vartheta)=a(\bs{r})\cos(\vartheta),\\
      a(\bs{r}):=\beta m\left[E(\bs{r})-E^\text{aux}(\bs{r})\right],
   \end{aligned}
   \label{eq:app1_2}
\end{align}
so that
\begin{align}
   \bar{\varrho}_0(\bs{r})=\zeta(\bs{r})\int\limits_0^{2\pi}\ud\varphi\int\limits_0^\pi \ud\vartheta\, \sin(\vartheta)\exp\left[a(\bs{r})\cos(\vartheta)\right]
                          =\zeta(\bs{r})4\pi\frac{\sinh[a(\bs{r})]}{a(\bs{r})},
\end{align}
which is equivalent to Eq.~(\ref{eq:ELE_rho0}).

For this orientation of the coordinate system, the normal component $P$ [Eq.~(\ref{eq:Pcomp})] of the polarization only depends
on the coefficient $f_{1,0}(\bs{r})$ of the expansion in Eq.~(\ref{eq:f}):
\begin{align}
   \begin{aligned}
      f_{1,0}(\bs{r})&=\int \ud^2\omega\, Y_{1,0}^\ast(\vartheta,\varphi) f(\bs{r},\bs{\omega})
                          =\sqrt{\frac{3}{4\pi}}\int \ud^2\omega\,\cos(\vartheta)\frac{\varrho_0(\bs{r},\bs{\omega})}{\bar{\varrho}_0(\bs{r})}\\
                     &=\sqrt{\frac{3}{4\pi}}\frac{a(\bs{r})}{4\pi\sinh[a(\bs{r})]}\int \ud^2\omega\,\cos(\vartheta)
                         \exp\left\{\beta m\bs{\omega}\cdot\left[\bs{E}(\bs{r})-\bs{E}^\text{aux}(\bs{r})\right]\right\}
   \end{aligned}
   \label{eq:app1_3}
\end{align}
where the asterisk $^\ast$ denotes the complex conjugate. In order to carry out the angular integration in Eq.~(\ref{eq:app1_3}) the scalar product therein
is treated like in Eq.~(\ref{eq:app1_2}) such that one obtains
\begin{align}
   \begin{aligned}
      f_{1,0}(\bs{r})=\sqrt{\frac{3}{4\pi}}\frac{a(\bs{r})}{4\pi\sinh[a(\bs{r})]}\int\limits_0^{2\pi}\ud\varphi\int\limits_0^\pi \ud\vartheta\, \sin(\vartheta)
                              \cos(\vartheta)\exp\left[a(\bs{r})\cos(\vartheta)\right]
                     =\sqrt{\frac{3}{4\pi}}\left\{\coth[a(\bs{r})]-\frac{1}{a(\bs{r})}\right\}=\sqrt{\frac{3}{4\pi}}\mathcal{L}[a(\bs{r})],
   \end{aligned}
   \label{eq:app1_4}
\end{align}
which is equivalent to Eq.~(\ref{eq:ELE_f10}).

\section{\label{app:large_distances}Asymptotic behavior at large distances from the wall}
In Sec.~\ref{subsection:ELE} the \textit{full} ELEs are presented in Eqs.~(\ref{eq:ELE1a}) and (\ref{eq:ELE1b})
which provide the most general description of the model.
From them the relevant reduced Eqs.~(\ref{eq:ELE1b}), (\ref{eq:ELE_rho0}), and (\ref{eq:ELE_f10})
are derived in Sec.~\ref{subsection:ELE}. They have to be solved numerically on a large but finite grid along the radial direction.
However, beyond the finite grid radial cutoff, the position of which is characterized by the length $r_g$
in this Appendix, assumptions concerning the profiles have to be made.
Here we focus on these assumptions
and on the resulting behavior at large distances from the wall,
where the external potentials [Eq.~(\ref{eq:V})] are identically zero and where it can be assumed that the quantities
\begin{align}
   \Delta\bs{E}(\bs{r})&:=\bs{E}(\bs{r})-\bs{E}^\text{aux}(\bs{r}),   \label{eq:DelE} \\
   \Delta\Phi_i(\bs{r})&:=\Phi(\bs{r})-\Phi_i^\text{aux}(\bs{r}), \quad i\in\{1,2\},
   \intertext{and}
   \Delta c_i^\text{hs,att}(\bs{r})&:=c_i^\text{hs}(\bs{r})-c_i^\text{hs,b}+c_i^\text{att}(\bs{r})-c_i^\text{att,b}, \quad i\in\{0,1,2\},   \label{eq:Delc}
\end{align}
are sufficiently small to allow for
linearization of the exponential functions in Eqs.~(\ref{eq:ELE1a}) and (\ref{eq:ELE1b}).
The following considerations focus on the general case of a spherical wall;
the corresponding results for a planar wall can be obtained from taking the limit of infinite wall radius.
With this, at large distances from the wall the ELEs~(\ref{eq:ELE1a}) and (\ref{eq:ELE1b}) take the form
\begin{align}
   \varrho_0&(\bs{r},\bs{\omega})\Big|_{|\bs{r}|>r_g}\simeq\frac{\varrho_0^\text{b}}{4\pi}\left[1+\Delta c_0^\text{hs,att}(\bs{r})+\beta m\bs{\omega}\cdot\Delta\bs{E}(\bs{r})\right],   \label{eq:ELE1a_lin}
\intertext{and}
   \varrho_i&(\bs{r})\Big|_{|\bs{r}|>r_g}\simeq\varrho_i^\text{b}\left[1+\Delta c_i^\text{hs,att}(\bs{r})-\beta q_i\Delta\Phi_i(\bs{r})\right],\quad i\in\{1,2\}.   \label{eq:ELE1b_lin}
\end{align}
Within this Appendix the focus is on the spatial decay of the electrostatic quantities $\Delta\Phi_i(\bs{r})$ and $\Delta\bs{E}(\bs{r})$
which follow from Eqs.~(\ref{eq:E})--(\ref{eq:Phiaux}). These expressions are independent of the contribution
$\Delta c_0^\text{hs,att}(\bs{r})$. The latter can enter $\Delta\Phi_i(\bs{r})$ and $\Delta\bs{E}(\bs{r})$ only
via the polarization $\bs{P}(\bs{r})$.
However, due to the angular integration in Eq.~(\ref{eq:P}), terms in $\varrho_0(\bs{r},\bs{\omega})$ which are independent of
the orientation $\bs{\omega}$, such as $\Delta c_0^\text{hs,att}(\bs{r})$ in Eq.~(\ref{eq:ELE1a_lin}), do not contribute to $\bs{P}(\bs{r})$.
In the limit of equal radii of the ions, $r_1=r_2$, the contributions $\Delta c_i^\text{hs,att}(\bs{r})$, $i\in\{1,2\}$, are equal
and  drop out of the sums in the first terms of Eqs.~(\ref{eq:E})--(\ref{eq:Phiaux}) due to the same absolute value of the ionic charges $q_1=-q_2$.
In this limit, as it will turn out in the following, the electrostatic contributions $\Delta\bs{E}(\bs{r})$ and
$\Delta\Phi_i(\bs{r})$ exhibit the same decay behavior.
The corresponding length scale is set by the Debye length $1/\kappa$ [Eqs.~(\ref{eq:mPB}) and
(\ref{eq:kappa})].
In Appendix~\ref{app:c} the asymptotic decay behavior of 
$\Delta c_i^\text{hs,att}(\bs{r})$ is discussed. There it is shown that in the limit of equal particle radii
the one-point direct correlation functions decay on the length scale of the
bulk correlation length $\xi$ emerging due to the presence of the hard spherical and attractive interactions.
Except very close to critical points,
at which $\xi$ diverges, $\xi$ is typically much smaller than $1/\kappa$ such that the contributions $\Delta c_i^\text{hs,att}(\bs{r})$
decay rapidly towards zero and can be neglected in Eqs.~(\ref{eq:ELE1a_lin}) and (\ref{eq:ELE1b_lin}) for large distances from the wall.
This can be expected to hold also for small differences between the particle radii which is the limit
we focus on in the present study.
Accordingly, our approximation for the ELEs at large distances from the wall reads
\begin{align}
   \varrho_0&(\bs{r},\bs{\omega})\Big|_{|\bs{r}|>r_g}\simeq\frac{\varrho_0^\text{b}}{4\pi}\left[1+\beta m\bs{\omega}\cdot\Delta\bs{E}(\bs{r})\right]   \label{eq:ELE1a_approx}
   \intertext{and}
   \varrho_i&(\bs{r})\Big|_{|\bs{r}|>r_g}\simeq\varrho_i^\text{b}\left[1-\beta q_i\Delta\Phi_i(\bs{r})\right],\quad i\in\{1,2\}.   \label{eq:ELE1b_approx}
\end{align}
Equations~(\ref{eq:E})--(\ref{eq:Phiaux}) are discussed in detail now. For this purpose the original integration domain $\mathbb{R}^3\backslash\mathcal{V}$
is extended to $\mathbb{R}^3$. Note that within $\mathcal{V}$ the integrands are identically zero for $r_1=r_2$. The extended domain
$\mathbb{R}^3$ is split into a region $|\bs{r}|\leq r_g$ close to the wall, where the full solutions are known,
and into a region $|\bs{r}|>r_g$ further away from the wall, 
where the number densities are approximated either as in Eqs.~(\ref{eq:ELE1a_lin}) and (\ref{eq:ELE1b_lin}) in the limit of equal ionic radii, or as in
Eqs.~(\ref{eq:ELE1a_approx}) and (\ref{eq:ELE1b_approx}) under the assumption of negligible one-point direct correlation function
differences. This leads to
\begin{align}
   \begin{aligned}
      \Delta E_\alpha(\bs{r})\simeq h\Bigg[-\sum\limits_{i=1}^2\beta q_i^2\varrho_i^\text{b}\int\limits_{\mathbb{R}^3} \ud^3r'\,K_{\alpha i}^{(1)}(\bs{d})\Delta\Phi_i(\bs{r'})
                                            -\frac{\beta m^2\varrho_0^\text{b}}{3}\sum\limits_{\gamma=x}^z\,\int\limits_{\mathbb{R}^3} \ud^3r'\,K_{\alpha\gamma}^{(2)}(\bs{d})\Delta E_\gamma(\bs{r'})+S_\alpha^{(1)}(\bs{r})\Bigg], \quad\alpha\in\{x,y,z\},
      \label{eq:DeltaE}
   \end{aligned}
   \intertext{and}
   \begin{aligned}
      \Delta\Phi_i(\bs{r})\simeq h\Bigg[-\sum\limits_{j=1}^2\beta q_j^2\varrho_j^\text{b}\int\limits_{\mathbb{R}^3} \ud^3r'\,K_{ij}^{(0)}(\bs{d})\Delta\Phi_j(\bs{r'})
                                         +\frac{\beta m^2\varrho_0^\text{b}}{3}\sum\limits_{\alpha=x}^z\,\int\limits_{\mathbb{R}^3} \ud^3r'\,K_{\alpha i}^{(1)}(\bs{d})\Delta E_\alpha(\bs{r'})+S_i^{(0)}(\bs{r})\Bigg], \quad i\in\{1,2\},
      \label{eq:DeltaPhi}
   \end{aligned}
\end{align}
with
\begin{align}
   K_{ij}^{(0)}(\bs{r})&:=\frac{1}{|\bs{r}|}\Theta[|\bs{r}|-(r_i+r_j)], &i,j\in\{1,2\},\\
   K_{\alpha i}^{(1)}(\bs{r})&:=\frac{r_\alpha}{|\bs{r}|^3}\Theta[|\bs{r}|-(r_0+r_i)], &\alpha\in\{x,y,z\},\;i\in\{1,2\}, \label{eq:K1}\\
   K_{\alpha\gamma}^{(2)}(\bs{r})&:=\left(\frac{\delta_{\alpha\gamma}}{|\bs{r}|^3}-\frac{3r_\alpha r_\gamma}{|\bs{r}|^5}\right)\Theta(|\bs{r}|-2r_0), &\alpha,\gamma\in\{x,y,z\},
\end{align}
\begin{align}
   \begin{aligned}
      S_i^{(0)}(\bs{r}):=&\sum\limits_{j=1}^2\;\int\limits_{|\bs{r'}|\leq r_g}\ud^3r'\frac{q_j}{|\bs{d}|}\left\{\left[\varrho_j(\bs{r'})-\varrho_j^\text{b}\right]+\beta q_j\varrho_j^\text{b}\Delta\Phi_j(\bs{r'})\right\}\Theta[|\bs{d}|-(r_i+r_j)]\\
          &+\int\limits_{|\bs{r'}|\leq r_g}\ud^3r'\left[\frac{\bs{P}(\bs{r'})\cdot\bs{d}}{|\bs{d}|^3}-\frac{1}{3}\varrho_0^\text{b}\beta m^2\frac{\Delta\bs{E}(\bs{r'})\cdot\bs{d}}{|\bs{d}|^3}\right]\Theta[|\bs{d}|-(r_0+r_i)]\\
          &+\int\limits_\mathcal{A}\ud^2r'\frac{\sigma}{|\bs{d}|}\Theta(|\bs{d}|-r_i), \quad i\in\{1,2\},\quad\bs{d}=\bs{r}-\bs{r'},
   \end{aligned}
\end{align}
\begin{align}
   \begin{aligned}
      S_\alpha^{(1)}(\bs{r}):=&\sum\limits_{i=1}^2\;\int\limits_{|\bs{r'}|\leq r_g}\ud^3r'\,q_i\frac{d_\alpha}{|\bs{d}|^3}\left\{\left[\varrho_i(\bs{r'})-\varrho_i^\text{b}\right]+\beta q_i\varrho_i^\text{b}\Delta\Phi_i(\bs{r'})\right\}\Theta[|\bs{d}|-(r_0+r_i)]\\
         &-\int\limits_{|\bs{r'}|\leq r_g}\ud^3r'\bigg\{\left[\frac{P_\alpha(\bs{r'})}{|\bs{d}|^3}-\frac{3d_\alpha\bs{P}(\bs{r'})\cdot\bs{d}}{|\bs{d}|^5}\right]
           -\frac{1}{3}\varrho_0^\text{b}\beta m^2\left[\frac{\Delta E_\alpha(\bs{r'})}{|\bs{d}|^3}-\frac{3d_\alpha\Delta\bs{E}(\bs{r'})\cdot\bs{d}}{|\bs{d}|^5}\right]\bigg\}\Theta(|\bs{d}|-2r_0)\\
         &+\int\limits_\mathcal{A}\ud^2r'\,\sigma\frac{d_\alpha}{|\bs{d}|^3}\Theta(|\bs{d}|-r_0), \quad \alpha\in\{x,y,z\},
   \end{aligned}
   \label{eq:S_alpha}
\end{align}
and
\begin{align}
   h&:=\frac{1}{4\pi\epsilon_0\epsilon_\text{ex}}. \label{eq:h}
\end{align}
In the limit $r_1=r_2$ the expressions of $S_i^{(0)}$ and $S_\alpha^{(1)}$ are odd functions of $\sigma$.
That is, for small surface charge densities $\sigma\rightarrow0$ both contributions are of the order $O(\sigma)$.
We will refer to this property at the end of this Appendix.
Note that $\delta_{\alpha\gamma}$ is the Kronecker symbol and the abbreviation $\bs{d}=\bs{r}-\bs{r'}$ for the spatial offset is still valid.
The Greek indices $\alpha,\gamma$ denote the vector components $x,y,z$ whereas the Latin indices $i,j$
denote the ion species $1,2$.
For example, the first term of Eq.~(\ref{eq:DeltaE}) can be derived as follows:
from Eqs.~(\ref{eq:E}) and (\ref{eq:Eaux}) one obtains
\begin{align}
    \bs{E}(\bs{r})&-\bs{E}^\text{aux}(\bs{r})=\Delta\bs{E}(\bs{r})\\
      &=\sum\limits_{i=1}^2\;\int\limits_{\mathbb{R}^3} \ud^3r'\frac{q_i}{4\pi\epsilon_0\epsilon_\text{ex}}\frac{\bs{d}}{|\bs{d}|^3}
         \left[\varrho_i(\bs{r'})-\varrho_i^\text{b}\right][1-\Theta(r_0+r_i-|\bs{d}|)]+\hdots \notag\\
      &=\frac{1}{4\pi\epsilon_0\epsilon_\text{ex}}\sum\limits_{i=1}^2\;\int\limits_{\mathbb{R}^3} \ud^3r'\,q_i\frac{\bs{d}}{|\bs{d}|^3}\left[\varrho_i(\bs{r'})-\varrho_i^\text{b}\right]\Theta[|\bs{d}|-(r_0+r_i)]+\hdots\;. \notag
\end{align}
In the next step, the asymptotic expression for the density profiles at large distances $|\bs{r}|>r_g$ from the wall [Eq.~(\ref{eq:ELE1b_lin}) or (\ref{eq:ELE1b_approx})] is used.
We note that $q_1\varrho_1^\text{b}+q_2\varrho_2^\text{b}=0$ due to local charge neutrality. The equation is written in terms of each component $\alpha\in\{x,y,z\}$ of the electric field:
\begin{align}
   \begin{aligned}
   \Delta E_\alpha(\bs{r})&\simeq\frac{1}{4\pi\epsilon_0\epsilon_\text{ex}}\Bigg\{\sum\limits_{i=1}^2\,\int\limits_{|\bs{r'}|\leq r_g}\ud^3r'\,q_i\frac{d_\alpha}{|\bs{d}|^3}\Big\{\left[\varrho_i(\bs{r'})-\varrho_i^\text{b}\right]
          -\varrho_i^\text{b}\left[\Delta c_i^\text{hs,att}(\bs{r'})-\beta q_i\Delta\Phi_i(\bs{r'})\right]\Big\}\Theta[|\bs{d}|-(r_0+r_i)]\\
        &+\sum\limits_{i=1}^2\;\int\limits_{\mathbb{R}^3} \ud^3r'\, q_i\frac{d_\alpha}{|\bs{d}|^3}\varrho_i^\text{b}\left[\Delta c_i^\text{hs,att}(\bs{r'})-\beta q_i\Delta\Phi_i(\bs{r'})\right]\Theta[|\bs{d}|-(r_0+r_i)]+\hdots\Bigg\}.
   \end{aligned}
\end{align}
In the limit of equal radii for the ions the contribution $\Delta c_i^\text{hs,att}(\bs{r'})$ drops out of the sum. The integrations with respect to positions
$|\bs{r'}|\leq r_g$ close to the wall are collected in $S_\alpha^{(1)}$ [Eq.~(\ref{eq:S_alpha})] and will not be considered further in this example.
Finally by using Eqs.~(\ref{eq:K1}) and (\ref{eq:h}) one has
\begin{align}
   \begin{aligned}
      \Delta E_\alpha(\bs{r})&\simeq\frac{1}{4\pi\epsilon_0\epsilon_\text{ex}}\Bigg[-\sum\limits_{i=1}^2\beta q_i^2\varrho_i^\text{b}
                 \int\limits_{\mathbb{R}^3} \ud^3r'\frac{d_\alpha}{|\bs{d}|^3}\Theta[|\bs{d}|-(r_0+r_i)]\Delta\Phi_i(\bs{r'})+\hdots\Bigg]\\
               &=h\Bigg[-\sum\limits_{i=1}^2\beta q_i^2\varrho_i^\text{b}\int\limits_{\mathbb{R}^3} \ud^3r'\,K_{\alpha i}^{(1)}(\bs{d})\Delta\Phi_i(\bs{r'})+\hdots\Bigg].
   \end{aligned}
\end{align}

In terms of the Fourier transform
\begin{align}
   \hat{f}(\bs{q}):=\int\limits_{\mathbb{R}^3} \ud^3r\, f(\bs{r})\exp(-\text{i}\bs{q}\cdot\bs{r})
   \label{eq:Fourier}
\end{align}
Eqs.~(\ref{eq:DeltaE}) and (\ref{eq:DeltaPhi}) lead to
a system of five linear equations for the components $\Delta\hat{E}_\alpha(\bs{q})$, $\alpha\in\{x,y,z\}$, of the vector $\Delta\hat{\bs{E}}(\bs{q})$ and
for $\Delta\hat{\Phi}_i(\bs{q})$ with $i\in\{1,2\}$:
\begin{align}
   \Delta\hat{E}_\alpha(\bs{q})&=h\left[-\sum\limits_{i=1}^2\beta q_i^2\varrho_i^\text{b} \hat{K}_{\alpha i}^{(1)}(\bs{q})\Delta\hat{\Phi}_i(\bs{q})
         -\frac{\beta m^2\varrho_0^\text{b}}{3}\sum\limits_{\gamma=x}^z\hat{K}_{\alpha \gamma}^{(2)}(\bs{q})\Delta \hat{E}_\gamma(\bs{q})+\hat{S}_\alpha^{(1)}(\bs{q})\right] \label{eq:SLE1}
\end{align}
and
\begin{align}
   \Delta\hat{\Phi}_i(\bs{q})&=h\left[-\sum\limits_{j=1}^2\beta q_j^2\varrho_j^\text{b}\hat{K}_{ij}^{(0)}(\bs{q})\Delta\hat{\Phi}_j(\bs{q})
         +\frac{\beta m^2\varrho_0^\text{b}}{3}\sum\limits_{\alpha=x}^z\hat{K}_{\alpha i}^{(1)}(\bs{q})\Delta\hat{E}_\alpha(\bs{q})+\hat{S}_i^{(0)}(\bs{q})\right]. \label{eq:SLE2}
\end{align}
With the 5-component vectors $\bs{v}$ and $\bs{s}$ as well as the $5\times5$ matrix $\bs{M}$,
\begin{align}
   \bs{v}=
   \begin{pmatrix}
      v_x \\
      v_y \\
      v_z \\
      v_1 \\
      v_2
   \end{pmatrix},\quad
   \bs{s}=
   \begin{pmatrix}
      s_x \\
      s_y \\
      s_z \\
      s_1 \\
      s_2
   \end{pmatrix},\quad
   \bs{M}=
   \begin{pmatrix}
      M_{xx} & M_{xy} & M_{xz} & M_{x1} & M_{x2} \\
      M_{yx} & M_{yy} & M_{yz} & M_{y1} & M_{y2} \\
      M_{zx} & M_{zy} & M_{zz} & M_{z1} & M_{z2} \\
      M_{1x} & M_{1y} & M_{1z} & M_{11} & M_{12} \\
      M_{2x} & M_{2y} & M_{2z} & M_{21} & M_{22}
   \end{pmatrix},
\end{align}
defined by
\begin{align}
   v_\alpha&:=\sqrt{\frac{\beta m^2\varrho_0^\text{b}}{3}}\;\Delta\hat{E}_\alpha(\bs{q}), &\alpha\in\{x,y,z\},\\
   v_i&:=\sqrt{\beta q_i^2\varrho_i^\text{b}}\;\Delta\hat{\Phi}_i(\bs{q}), &i\in\{1,2\},\\
   s_\alpha&:=h\sqrt{\frac{\beta m^2\varrho_0^\text{b}}{3}}\;\hat{S}_\alpha^{(1)}(\bs{q}), &\alpha\in\{x,y,z\},\\
   s_i&:=h\sqrt{\beta q_i^2\varrho_i^\text{b}}\;\hat{S}_i^{(0)}(\bs{q}), &i\in\{1,2\},\\
   M_{\alpha\gamma}&:=h\frac{\beta m^2\varrho_0^\text{b}}{3}\hat{K}_{\alpha\gamma}^{(2)}(\bs{q}), &\alpha,\gamma\in\{x,y,z\},\\
   M_{\alpha i}&:=-M_{i \alpha}:=h\sqrt{\frac{\beta m^2\varrho_0^\text{b}}{3}\beta q_i^2\varrho_i^\text{b}}\;\hat{K}_{\alpha i}^{(1)}(\bs{q}), &\alpha\in\{x,y,z\},\;i\in\{1,2\},\\
   \intertext{and}
   M_{ij}&:=h\sqrt{\beta q_i^2\varrho_i^\text{b} \beta q_j^2\varrho_j^\text{b}}\;\hat{K}_{ij}^{(0)}(\bs{q}), &i,j\in\{1,2\},
\end{align}
the system of linear equations~(\ref{eq:SLE1}) and (\ref{eq:SLE2}) has the form
\begin{align}
   (\bs{1}+\bs{M})\cdot\bs{v}=\bs{s} \Leftrightarrow \bs{v}=(\bs{1}+\bs{M})^{-1}\cdot\bs{s}
   \label{eq:Yvon}
\end{align}
with the identity matrix $\bs{1}$.
With the abbreviations
\begin{align}
   k^2&:=8\pi\beta e^2hI,\\
   p^2&:=\beta m^2h\varrho_0^\text{b},\\
   \bar{r}_i&:=\frac{r_i}{r_0},\quad i\in\{1,2\},\\
   \bar{k}&:=kr_0,\\
   x&:=q r_0:=\sqrt{\bs{q}\cdot\bs{q}}\,r_0,\\
   a&:=\frac{\pi}{6}p^2\;\frac{\sin(2x)-2x\cos(2x)}{x^3},\\
   b_j&:=-\text{i}\sqrt{\frac{2\pi}{3}}p\bar{k}\;\frac{\sin[x(1+\bar{r}_j)]}{x^2(1+\bar{r}_j)},\quad j\in\{1,2\},\\
   \intertext{and}
   M_{ij}&=\frac{\bar{k}^2}{2}\frac{\cos[x(\bar{r}_i+\bar{r}_j)]}{x^2},\quad i,j\in\{1,2\},
\end{align}
one obtains
\begin{align}
   \begin{aligned}
   \bs{1}+\bs{M}=
      \begin{pmatrix}
         1-a   &     0     &     0     &     0     &     0     \\
         0     &    1-a    &     0     &     0     &     0     \\
         0     &     0     &    1+2a   &    b_1    &    b_2    \\
         0     &     0     &    -b_1   &  1+M_{11} &  M_{12}   \\
         0     &     0     &    -b_2   &   M_{12}  &  1+M_{22}
      \end{pmatrix}.
   \end{aligned}
\end{align}
For various combinations of the indices $\alpha,i$ and $\alpha,\gamma$ the Fourier transforms
$\hat{K}_{\alpha i}^{(1)}(\bs{q})$ and $\hat{K}_{\alpha\gamma}^{(2)}(\bs{q})$ are identically zero.
Therefore the matrix $\bs{1}+\bs{M}$ exhibits this kind of block structure.
Equation~(\ref{eq:Yvon}) relates the response of the fields $\bs{v}$ to an external perturbation $\bs{s}$, i.e., it
is a multidimensional analogue of Yvon's equation in Fourier space \cite{Hansen1976}. Therefore, up to irrelevant factors, the
matrix $\hat{\bs{G}}(q):=(\bs{1}+\bs{M})^{-1}$ is the Fourier transform of the bulk two-point direct correlation functions
of the field components $\Delta E_{x,y,z}$ and of the potential differences $\Delta\Phi_{1,2}$.
With the inverse Fourier transform
\begin{align}
   \bs{G}(r):=\frac{1}{4\pi^2\text{i}r}\int\limits_{-\infty}^\infty \ud q\,q\,\hat{\bs{G}}(q)\exp(\text{i}qr),\quad r:=|\bs{r}|,
   \label{eq:inverse_Fourier}
\end{align}
one obtains $\bs{G}(r)$, which determines the asymptotic spatial dependence of
$\Delta E_{x,y,z}(\bs{r})$ and $\Delta\Phi_{1,2}(\bs{r})$.
The integral in Eq.~(\ref{eq:inverse_Fourier}) can be studied by using the residue theorem.
As a consequence the exponential decay of $\bs{G}(r)$ is determined by the poles of $\hat{\bs{G}}(q)$ \cite{Evans1993,Evans1994}.
The pole $q'+\text{i}q''\in\mathbb{C}$, with $q',q''\in\mathbb{R}$, of $\hat{\bs{G}}(q)$
with the smallest imaginary part $|q''|$ determines the asymptotic decays of $\Delta E_{x,y,z}(\bs{r})$
and $\Delta\Phi_{1,2}(\bs{r})$ on the length scale $1/|q''|$ away from the charged wall.
Since, according to Cramer's rule, the inverse matrix $(\bs{1}+\bs{M})^{-1}=\hat{\bs{G}}(q)\propto 1/\text{det}(\bs{1}+\bs{M})$
is proportional to the reciprocal of $\text{det}(\bs{1}+\bs{M})$,
the poles of $\hat{\bs{G}}(q)$ are given by the roots of the determinant $\text{det}(\bs{1}+\bs{M})$:
\begin{align}
   \begin{aligned}
      \text{det}(\bs{1}+\bs{M})=(1-a)^2[(1+2a)(1+M_{11})(1+M_{22})-2b_1b_2M_{12}+b_2^2(1+M_{11})
                                     -M_{12}^2(1+2a)+b_1^2(1+M_{22})]=0.
   \end{aligned}
   \label{eq:det}
\end{align}
Equation~(\ref{eq:det}) can be solved numerically.  We find for our parameter choices (see Sec.~\ref{subsec:parameters}) purely imaginary
roots $\text{i}q''$, i.e., the asymptotic decay of $\Delta E_{x,y,z}(\bs{r})$ and $\Delta\Phi_{1,2}(\bs{r})$ is monotonic.
However, the important finding is that all electric field components
and electric potentials decay on the \textit{same} length scale $1/|q''|$ and are proportional to $\sigma$
in the limit of equal radii $r_1=r_2$ of the ions and for $\sigma\rightarrow0$.
This finding is relevant for Sec.~\ref{subsec:far_from_wall} and in particular for Eq.~(\ref{eq:constants}).

\section{\label{app:c}Asymptotic decay of the one-point direct correlation functions}
Within this Appendix the asymptotic decay behavior of the one-point direct correlation functions $c_{0,1,2}^\text{hs}(\bs{r})$ and $c_{0,1,2}^\text{att}(\bs{r})$
is examined at positions far away from the wall. This behavior is related to the decay behavior of the number densities $\bar{\varrho}_0(\bs{r})$
and $\varrho_{1,2}(\bs{r})$ which fulfill the ELEs~(\ref{eq:ELE_rho0}) and (\ref{eq:ELE1b}).
We introduce the deviations of the number densities from their respective bulk values,
$\Delta\varrho_0(\bs{r}):=\bar{\varrho}_0(\bs{r})-\varrho_0^\text{b}$ and $\Delta\varrho_{1,2}(\bs{r}):=\varrho_{1,2}(\bs{r})-\varrho_{1,2}^\text{b}$,
and use the notation in Eqs.~(\ref{eq:DelE})--(\ref{eq:Delc}) in order to rewrite Eqs.~(\ref{eq:ELE_rho0}) and (\ref{eq:ELE1b}) as
\begin{align}
   \ln\left[1+\frac{\Delta\varrho_0(\bs{r})}{\varrho_0^\text{b}}\right]+\beta V_0(\bs{r})-\Delta c_0^\text{hs,att}(\bs{r})
        -\ln\left\{\frac{\sinh[\beta m \Delta E(\bs{r})]}{\beta m \Delta E(\bs{r})}\right\}=0	\label{eq:appc_0'}
   \intertext{and, $i\in\{1,2\}$,}
   \ln\left[1+\frac{\Delta\varrho_i(\bs{r})}{\varrho_i^\text{b}}\right]+\beta V_i(\bs{r})-\Delta c_i^\text{hs,att}(\bs{r})
        +\beta q_i\Delta\Phi_i(\bs{r})=0.   \label{eq:appc_1'_2'}
\end{align}
Within this Appendix, a spherical wall is discussed, as in Appendix~\ref{app:large_distances}, and we consider
the case that all particle species have the same radius: $r_0=r_1=r_2$.
As a consequence the one-point direct correlation functions and the differences of the electrostatic potentials are the same
for all species, i.e., $\Delta c_{0,1,2}^\text{hs,att}(\bs{r})=:\Delta c^\text{hs,att}(\bs{r})$ and $\Delta\Phi_{1,2}(\bs{r})=:\Delta\Phi(\bs{r})$.
In the following the equation for the solvent [Eq.~(\ref{eq:appc_0'})] is discussed in detail so that the presented procedure
can be applied analogously to the two equations for the ionic species in Eq.~(\ref{eq:appc_1'_2'}).
The Fourier transform [Eq.~(\ref{eq:Fourier})] of Eq.~(\ref{eq:appc_0'}) is given by
\begin{align}
   \int\limits_{\mathbb{R}^3} \ud^3r\,\exp(-\text{i}\bs{q}\cdot\bs{r})\left\{\ln\left[1+\frac{\Delta\varrho_0(\bs{r})}{\varrho_0^\text{b}}\right]+\beta V_0(\bs{r})-\Delta c^\text{hs,att}(\bs{r})
           -\ln\left\{\frac{\sinh[\beta m \Delta E(\bs{r})]}{\beta m \Delta E(\bs{r})}\right\}\right\}=0.
   \label{eq:appc_0''}
\end{align}
We introduce the length $r_g$ large enough such that
\begin{align}
   \beta V_{0,1,2}(\bs{r})\big|_{|\bs{r}|>r_g}&=0,   \label{eq:appc_approx_V}\\
   \frac{\sinh[\beta m \Delta E(\bs{r})]}{\beta m \Delta E(\bs{r})}\bigg|_{|\bs{r}|>r_g}&\simeq \frac{\beta m \Delta E(\bs{r})}{\beta m \Delta E(\bs{r})}=1,   \label{eq:appc_approx_E}\\
   \intertext{and, $i\in\{0,1,2\}$,}
   \ln\left[1+\frac{\Delta\varrho_i(\bs{r})}{\varrho_i^\text{b}}\right]\Bigg|_{|\bs{r}|>r_g}&\simeq\frac{\Delta\varrho_i(\bs{r})}{\varrho_i^\text{b}}.   \label{eq:appc_approx_rho}
\end{align}
The integration in Eq.~(\ref{eq:appc_0''}) over the whole space can be split into two
domains, one with $|\bs{r}|\leq r_g$, where the full ELE~(\ref{eq:appc_0'}) is integrated,
and another one with $|\bs{r}|>r_g$, where the ELE~(\ref{eq:appc_0'}) is approximated according to Eqs.~(\ref{eq:appc_approx_V})--(\ref{eq:appc_approx_rho}).
We make use of the fact that the equilibrium density $\bar{\varrho}_0(\bs{r})$
fulfills the ELE~(\ref{eq:appc_0'}) locally which is why the integral in the domain $|\bs{r}|\leq r_g$ vanishes. Therefore, with the approximations
in Eqs.~(\ref{eq:appc_approx_V})--(\ref{eq:appc_approx_rho}), the Fourier transform in Eq.~(\ref{eq:appc_0''}) reads
\begin{align}
   \int\limits_{|\bs{r}|>r_g}\ud^3r\,\exp(-\text{i}\bs{q}\cdot\bs{r})\left[\frac{\Delta\varrho_0(\bs{r})}{\varrho_0^\text{b}}-\Delta c^\text{hs,att}(\bs{r})\right]=0.
   \label{eq:appc_0'''}
\end{align}
The one-point direct correlation function $\Delta c^\text{hs,att}(\bs{r})$, which is a functional of the number densities $\varrho_{0,1,2}$ [see Eq.~(\ref{eq:c})],
can be expressed in terms of a Taylor series expansion with respect to the bulk value:
\begin{align}
   \Delta c^\text{hs,att}(\bs{r})\simeq\int\limits_{\mathbb{R}^3} \ud^3r'\, c^{(2)\text{hs,att,b}}(\bs{r}-\bs{r'})[\Delta\varrho_0(\bs{r'})+\Delta\varrho_1(\bs{r'})+\Delta\varrho_2(\bs{r'})].
   \label{eq:appc_functional_taylor}
\end{align}
Since the integral in Eq.~(\ref{eq:appc_0'''}) is restricted to positions far away from the wall, i.e., $|\bs{r}|>r_g$, where the number densities are close to their respective
bulk values and hence the deviations $|\Delta\varrho_{0,1,2}|$ are small,
in Eq.~(\ref{eq:appc_functional_taylor}) only terms up to and including linear order in $\Delta\varrho_{0,1,2}$ are taken into account.
Note that $\Delta c^\text{hs,att}$ [Eq.~(\ref{eq:Delc})] measures the difference of the one-point direct correlation functions
from their respective bulk values. Therefore evaluation in the bulk leads to $\Delta c^\text{hs,att}=0$ which is the zeroth order
of the expansion in Eq.~(\ref{eq:appc_functional_taylor}).
The quantity $c^{(2)\text{hs,att,b}}$ denotes the bulk two-point direct
correlation function governed by the hard spherical (hs) and the attractive (att) interaction, respectively.
In Eq.~(\ref{eq:appc_0'''}) a conveniently chosen zero is added such that the original integration over $|\bs{r}|>r_g$ is written
in terms of a Fourier integral and an integration over the domain $|\bs{r}|\leq r_g$ in order to obtain
\begin{align}
   \begin{aligned}
      \frac{\Delta\hat{\varrho}_0(\bs{q})}{\varrho_0^\text{b}}-\hat{c}^{(2)\text{hs,att,b}}(\bs{q})[\Delta\hat{\varrho}_0(\bs{q})+\Delta\hat{\varrho}_1(\bs{q})+\Delta\hat{\varrho}_2(\bs{q})]
              -F_0(\bs{q})=0,\\
      F_0(\bs{q}):=\int\limits_{|\bs{r}|\leq r_g}\ud^3r\,\exp(-\text{i}\bs{q}\cdot\bs{r})\left\{\frac{\Delta\varrho_0(\bs{r})}{\varrho_0^\text{b}}
                 -\int\limits_{\mathbb{R}^3} \ud^3r'\, c^{(2)\text{hs,att,b}}(\bs{r}-\bs{r'})\sum\limits_{i=0}^2\Delta\varrho_i(\bs{r'}) \right\}.
   \end{aligned}
   \label{eq:appc_0(5)}
\end{align}
The same procedure can be applied to the sum of the two equations in Eq.~(\ref{eq:appc_1'_2'}). The contribution of the electric potential drops out of the sum
because the ion species carry a charge of the same absolute value $q_1=-q_2$, thus leading to
\begin{align}
   \begin{aligned}
      &\frac{\Delta\hat{\varrho}_1(\bs{q})+\Delta\hat{\varrho}_2(\bs{q})}{I}-2\hat{c}^{(2)\text{hs,att,b}}(\bs{q})[\Delta\hat{\varrho}_0(\bs{q})+\Delta\hat{\varrho}_1(\bs{q})+\Delta\hat{\varrho}_2(\bs{q})]-F_{1,2}(\bs{q})=0,\\
      &F_{1,2}(\bs{q}):=\int\limits_{|\bs{r}|\leq r_g}\ud^3r\,\exp(-\text{i}\bs{q}\cdot\bs{r})\left\{\frac{\Delta\varrho_1(\bs{r})+\Delta\varrho_2(\bs{r})}{I}
                 -2\int\limits_{\mathbb{R}^3} \ud^3r'\, c^{(2)\text{hs,att,b}}(\bs{r}-\bs{r'})\sum\limits_{i=0}^2\Delta\varrho_i(\bs{r'}) \right\}.
   \end{aligned}
   \label{eq:appc_3'''}
\end{align}
The functions $F_0(\bs{q})$ and $F_{1,2}(\bs{q})$ are entire, i.e., they do not possess any poles in $\bs{q}\in\mathbb{C}^3$,
because the outer-most integrations
in Eqs.~(\ref{eq:appc_0(5)}) and (\ref{eq:appc_3'''}) are those of continuous integrands which are entire in $\bs{q}$ over a compact domain.
The sum of Eqs.~(\ref{eq:appc_0(5)}) and (\ref{eq:appc_3'''}),
\begin{align}
   \Delta\hat{\varrho}_0(\bs{q})+\Delta\hat{\varrho}_1(\bs{q})+\Delta\hat{\varrho}_2(\bs{q})=\frac{\varrho_0^\text{b}F_0(\bs{q})+IF_{1,2}(\bs{q})}{\displaystyle1-\hat{c}^{(2)\text{hs,att,b}}(\bs{q})(\varrho_0^\text{b}+2I)},
   \label{eq:appc_4}
\end{align}
is an analogue of Yvon's equation in Fourier space \cite{Hansen1976} because it relates the number densities
with an external perturbation given by the numerator on the right hand side of Eq.~(\ref{eq:appc_4}).
The expression $S(\bs{q}):=[1-\hat{c}^{(2)\text{hs,att,b}}(\bs{q})(\varrho_0^\text{b}+2I)]^{-1}$ is the
bulk structure factor \cite{Hansen1976}. Hence, following the line of argument in Appendix~\ref{app:large_distances}
and recognizing that the numerator in Eq.~(\ref{eq:appc_4}) does not have poles,
the asymptotic decay of the sum $\Delta\varrho_0(\bs{r})+\Delta\varrho_1(\bs{r})+\Delta\varrho_2(\bs{r})$ is given by the pole
$q'+\text{i}q''$ of $S(\bs{q})$ with the smallest imaginary part $|q''|$, which sets the length scale of the decay.
Here the length scale can be identified as the bulk correlation length $\xi=1/|q''|$ emerging from the hard spherical and attractive interactions.
Far away from the wall, i.e., at positions $\bs{r}$ at which the approximations in Eqs.~(\ref{eq:appc_approx_V})--(\ref{eq:appc_approx_rho}) can be applied,
the ELE for the solvent Eq.~(\ref{eq:appc_0'}) and the sum of the ELEs for the ions in Eq.~(\ref{eq:appc_1'_2'}) are given by
\begin{align}
   \frac{\Delta\varrho_0(\bs{r})}{\varrho_0^\text{b}}=\Delta c^\text{hs,att}(\bs{r})   \label{eq:appc_lin_ELE_solvent}
   \intertext{and}
   \frac{\Delta\varrho_1(\bs{r})}{\varrho_1^\text{b}}+\frac{\Delta\varrho_2(\bs{r})}{\varrho_2^\text{b}}=2\Delta c^\text{hs,att}(\bs{r}).   \label{eq:appc_lin_ELE_ions}
\end{align}
From Eqs.~(\ref{eq:appc_lin_ELE_solvent}) and (\ref{eq:appc_lin_ELE_ions}) it follows that the decay of the one-point direct correlation function difference
$\Delta c^\text{hs,att}(\bs{r})$ is given by the decay
of the number densities $\Delta\varrho_{0,1,2}(\bs{r})$. That is, $\Delta c^\text{hs,att}(\bs{r})$ decays on the length scale of the bulk correlation length $\xi$.
This result is used in Sec.~\ref{subsec:far_from_wall} and
in Appendix~\ref{app:large_distances} in order to justify the neglect of the one-point direct correlation function differences
in Eqs.~(\ref{eq:ELE1a_approx}) and (\ref{eq:ELE1b_approx}).

\end{widetext}


\end{document}